# How to enhance the applicability of Murnaghan equation of state

V.V. Bannikov

*Institute of Solid State Chemistry, Ural Branch of RAS, 620990, Ekaterinburg, Pervomayskaya st., 91, Russia*

ABSTRACT

The paper presents the brief overview of known isothermal equations of state for solids and of theoretical background to construct them, focusing on widely used Murnaghan equation. Since the latter has a limited range of application, failing to reproduce reliably both near-equilibrium and high-pressure regions simultaneously, the paper proposes simple, but rather efficient way to improve it. The suggested modification favors its applicability at high pressures, allowing to expand considerably the region of reliable reproducing of experimental or computational data. The proposed improved equation has been tested in fitting of $U$(V) data, obtained within *ab initio* computations, as well as of experimental V–$P$ data for a representative selection of solids. In comparison with ordinary Murnaghan equation, it has found significant advantages, extending the range, where the computational data may be reproduced trustworthy (from the equilibrium state to the pressures, appreciably higher, the ordinary equation could). As for the fitting of experimental data, the proposed improved Murnaghan equation reproduces them, in the presented examples, at least not worse than other well-known equations (Vinet, Birch-Murnaghan, Mao) do. Some aspects of fitting process, as well as some alternative approaches to obtain the proposed equation of state in various forms are also discussed.

## 1 Introduction

The equation of state (EOS) is the relation that establishes the interdependence between the internal physical characteristics of a substance (such as temperature, volume, magnetization, etc.) and the measures of external conditions, to which it is subjected (i.e. pressure, external fields, etc). It is obvious that knowing of explicit form of EOS for a particular substance is extremely important for lots of applications.

First, it makes the general (and rather abstract) thermodynamic theory to be able to describe quantitatively the properties and behavior of a substance. Further, the experimental measurements (or, alternatively, *ab initio* modeling) may require the explicit and plausible analytical relation between the physical quantities under interest. Knowing it would allow to fit experimental (or computational) data and to evaluate the internal properties of a substance (i.e. elastic characteristics from *P*-V-T measurements, averaged atomic magnetic moments from M-H-T measurements, etc.) in terms of EOS fitting parameters, or to predict its behavior under conditions, unavailable for current experimental investigations. It is unlikely to be possible, however, to propose a "flawlessly right" (and practically usable) EOS for a given substance (except some model systems, like ideal gas), based only on the fundamental thermodynamic consideration. There are various approaches to find the appropriate EOS for one or the other kind of materials: the empirical way, based on the generalization of available experimental data, the phenomenological approach, based on quite general physical regularities (such as the minimum of energy of solids near the equilibrium state), and the approach, based on the assumption of a particular microscopic model of inter-atomic interactions for a specified class of substances. Each of them has its own evident advantages and drawbacks, moreover, the behavior of the same substance can be described by different EOSs (obtained within various ways) at different conditions with different accuracy. To construct more advanced EOSs and to improve the known ones (i.e. to propose a way to enhance the region of their reliable applicability, avoiding unnecessary complications of their math forms) is an important objective of the theoretical background of materials science.

For condensed matter physics and chemistry, the isothermal EOS, describing volume *vs* hydrostatic pressure dependence (or *vice versa*) at constant temperature, and represented in $P=P(\mathrm{V})$ general form, are of particular interest. The condensed substance is assumed to be characterized by a finite equilibrium volume at zero pressure, and to be isotropic, i.e. its volume would be sufficient to characterize completely its geometry during considerations (in fact, the monocrystals of cubic symmetry behave under hydrostatic pressure as isotropic solids, so, this suggestion is applicable for them too). The fitting parameters of such EOS are supposed to be temperature-dependent, and, if temperature varies slowly, it can be applicable also for the description of thermodynamic processes in condenses substances. The development of different kinds of isothermal $P$=$P$(V) EOSs has been proceeding primarily in association with geophysical applications – to describe the behavior of substance at extremely high pressures – in Earth mantle (up to ~150 GPa) and core (~150–350 GPa), but their helpfulness is not limited to it. The high-pressure synthesis is the efficient and prospective technique in solid state chemistry, and it can be performed under the pressures just one order of magnitude less or even comparable with "geophysical" ones. For example, a series of lanthanum hydrides $LaH_{n+\delta}$ (n=3-10) have been synthesized at the pressure range 50-176 GPa [1], the platinum nitride and carbide have been obtained and investigated at 45-50 GPa and up to 85 GPa, respectively [2, 3], the oxide perovskite $BaVO_3$ was synthesized at 15 GPa [4], and so on. The EOS, able to describe robustly the behavior of the participants of reactions in wide pressure range (from ambient conditions up to mentioned values), would be the highly valuable tool for quantitative description and prediction the feasibility of such kinds of processes. Both for first-principles *total energy vs volume* computations and for direct V–$P$ measurements, the isothermal $P$=$P$(V) EOS, reliable in rather wide pressure range, also is of interest, – to treat the obtained results for the investigated substances and to evaluate their elastic characteristics. In order to survey the particular forms of the isothermal $P$=$P$(V) EOSs and their origin, as well as their comparative applicability to behavior of particular solids and liquids (water, mercury, rock salt, different oxides like MgO, $Mg_2SiO_4$, corundum, olivine, etc.) and to geophysical descriptions, the textbook O.L. Anderson [5], the old, but very comprehensive review paper of J.R. Macdonald [6], and the paper of F.D. Stacey with co-workers [7] can be recommended.

Though the great number of EOSs, proposed up to now, most of them involve the fitting parameters with the same physical meaning: the equilibrium volume $V_0$ (i.e. in absence of external pressure), the values of isothermal bulk modulus $B = -V\cdot(\partial P/\partial V)_T$ and of its pressure derivative $(\partial B/\partial P)_T$ at zero pressure, $B_0$ and $Bp$, respectively. The advanced forms of EOS may also include the second pressure derivative of bulk modulus, $Bpp = (\partial^2 B/\partial P^2)_{T,P=0}$, or some other parameters. We shall mention here only a few of known EOSs, most widely employed (according to author observations, at least) in the applications of solid state physics and chemistry, namely, the Murnaghan EOS [8]:

$$P^{(\mathrm{M})}(\mathrm{V}) = \frac{B_0}{Bp}\cdot\left[\left(\frac{\mathrm{V}_0}{\mathrm{V}}\right)^{Bp} - 1\right] \quad (1),$$

the Birch-Murnaghan EOS (of third order) [9]:

$$P^{(\mathrm{BM3})}(\mathrm{V}) = \frac{3B_0}{2}\cdot\left[\left(\frac{\mathrm{V}_0}{\mathrm{V}}\right)^{7/3} - \left(\frac{\mathrm{V}_0}{\mathrm{V}}\right)^{5/3}\right]\times\left\{1 + \frac{3}{4}(Bp-4)\cdot\left(\left(\frac{\mathrm{V}_0}{\mathrm{V}}\right)^{2/3} - 1\right)\right\} \quad (2),$$

the Vinet (or Vinet-Rose) EOS [10, 11]:

$$P^{(\mathrm{VR})}(\mathrm{V}) = 3B_0\cdot\left[\left(\frac{\mathrm{V}_0}{\mathrm{V}}\right)^{2/3} - \left(\frac{\mathrm{V}_0}{\mathrm{V}}\right)^{1/3}\right]\times exp\left(\frac{3}{2}(Bp-1)\cdot\left[1 - \left(\mathrm{V}_0/\mathrm{V}\right)^{-1/3}\right]\right) \quad (3),$$

and Anton-Schmidt EOS [12, 13]:

$$P^{(\mathrm{AS})}(\mathrm{V}) = B_0\cdot\left(\mathrm{V}/\mathrm{V}_0\right)^{n}\cdot\ln\left(\mathrm{V}_0/\mathrm{V}\right) \quad (4)$$

(the latter equation involves the dimensionless parameter $n$, which was associated in [12] with Gruneisen parameter $\gamma$ as $-\gamma = n + 1/6$, though it is straightforward to obtain that $n = -Bp/2$). Though the relatively simple form and just a few fitting parameters are the advantages of these and related EOSs, their common blind side is the lack of their "*flexibility*". It can be problematic for them to reproduce the experimental or computational V-$P$ data equally reliably both for high pressure region and near zero pressure *simultaneously*., i.e. in general case, the excellent quality of fitting near equilibrium state may result in poor fitting at high pressures and *vice versa*. Undoubtedly, the task to improve the "flexibility" of EOSs, i.e. to extend their reliable fitting region is very important for their practical applications, and manifold attempts in this field have been undertaken (resulting in the variety of different kinds of EOSs). On the other hand, the corrected EOSs often become more complicated, involve more fitting parameters and do not guarantee yet the better fitting quality and universality (see the further section), so, the ideas, concerning the effective modification of different EOSs, are still welcome. The present paper proposes the simple, but rather elegant and efficient approach (in the modest author`s opinion) to improve one of the most widely known EOSs in solid state physics – the Murnaghan EOS (1). The suggested modified equation was applied to fit the computational *total energy vs volume* data for the representative sampling of solids with cubic crystalline structure, and the results exhibited the significant enlargement of the reliable fitting region, in other words, the applicability in considerably wider region of pressure values – in comparison with the original EOS. Being applied to some recently published *volume vs pressure* experimental data, the proposed EOS has revealed at least not worse quality of fitting, than the known EOSs can ensure. Therewith, the modified equation requires just one additional fitting parameter, and its form is not too complicated, as compared with (1).

The article is organized as follows. The "Theoretical overview" section discusses in details the possible approaches to constructing the EOSs, the related problems and possible ways to solve them, reviewing along the way some more widely known EOSs (Sub-section 1.1, a reader, well-experienced in this field, can skip it on first reading). Further, it regards the mentioned problem of "flexibility" of EOSs and makes some hints, how it is possible to improve it (Sub-section 1.2). "The improved Murnaghan EOS" section proposes the certain way to improve the Murnaghan EOS (1), making it applicable at higher pressures. The next sections are devoted to the application of improved Murnaghan EOS to fit both the available computational and experimental data – in comparison with some other EOSs, and discuss also some aspects of the fitting process, as well as some features of the proposed improved EOS and a possible alternative approach to derive it in somewhat different form.

# 2 The theoretical overview

## 2.1 The ways to construct EOS

*(This sub-section can be skipped on first reading, going straight to Sub-section 1.2)*

For solids, the widely known phenomenological way to construct EOSs is the formal expansion of free Helmholtz energy $F$ into Maclaurin series in terms of small dimensionless parameter $f$ (see [5, 7]), i.e.:

$$F = A_0 + A_2{\cdot}f^2 + A_3{\cdot}f^3 + A_4{\cdot}f^4 + A_5{\cdot}f^5 + A_6{\cdot}f^6 + \dots , \quad (5)$$

where $f = f(V)$ tends to zero as V tends to equilibrium value $V_0$, and the coefficients $A_i$ are supposed to be the temperature-dependent quantities. At constant temperature $F$ has a minimum at $V=V_0$, so the linear term in (5) is zero, and $A_2>0$. For instance, the Lagrangian or Eulerian finite strains for a simple isotropic hydrostatic compression, i.e.:

$$\varepsilon_L = [(V/V_0)^{2/3} - 1]/2, \text{ or } \varepsilon_E = [1 - (V/V_0)^{-2/3}]/2, \quad (6)$$

respectively, can be regarded as $f$. Taking $f=-\varepsilon_E$ and taking into account that $P = -(\partial F/\partial V)_T = -(\partial F/\partial f)_T{\cdot}(df/dV)$, we can truncate the series (5) to the second, third or fourth term and obtain the Birch-Murnaghan EOS of the second:

$$P^{(BM2)}(V) = \frac{3B_0}{2}\cdot\left[\left(\frac{V_0}{V}\right)^{7/3} - \left(\frac{V_0}{V}\right)^{5/3}\right] \quad (7)$$

third (2) or fourth order:

$$P^{(BM4)}(V) = \frac{3B_0}{2}\cdot\left[\left(\frac{V_0}{V}\right)^{7/3} - \left(\frac{V_0}{V}\right)^{5/3}\right]\times$$
$$\times\left\{1+\frac{3}{4}(Bp-4)\cdot\left(\left(\frac{V_0}{V}\right)^{2/3}-1\right)+\frac{1}{24}(9\cdot B_0\cdot Bpp - 9\cdot Bp\cdot(7-Bp)+143)\cdot\left(\left(\frac{V_0}{V}\right)^{2/3}-1\right)^2\right\} \quad (8)$$

respectively. To derive them we should take into account that:

$$B = -V{\cdot}(\partial P/\partial V)_T,\ (\partial B/\partial P)_T = -(V/B){\cdot}(\partial B/\partial V)_T, \text{ and}$$
$$(\partial^2 B/\partial P^2)_T = (V/B^2){\cdot}\{[1-(V/B)(\partial B/\partial V)_T](\partial B/\partial V)_T + V{\cdot}(\partial^2 B/\partial V^2)_T\}, \quad (9)$$

it allows to evaluate $B_0$, $Bp$ and $Bpp$ parameters, as they were introduced above, and to express the $A_i$ constants in (5) in terms of them. During the formal considerations, the expressions (6) are not unique for strains, for instance, they may be modified replacing ±2/3 powers with another ones, the EOSs, analogously constructed on the base of such trial strains, have been considered and compared with each other in [14]. Alternatively, the Hencky strain $\varepsilon_H=(1/3)\cdot ln(V/V_0)$ can be regarded, and, taking $f=\varepsilon_H$, so-called *logarithmic* EOS [15] can be derived in a similar way:

$$P^{(\log)}(V) = \frac{B_0 \cdot V_0}{V} \cdot \ln\left(V_0/V\right) \cdot \left[1 + \frac{(Bp-2)}{2} \cdot \ln\left(V_0/V\right)\right] \quad (10)$$

(confining with the third order). Moreover, during the formal considerations, it is not necessary to demand a particular physical meaning for $f(V)$ function, it is sufficient for it just to satisfy the requirements, specified above. It can be chosen in manifold ways (including, of course, the simple relative compression $\delta = (V-V_0)/V_0$ too), hence, the wide variety of different kinds of EOS can be constructed in such a manner. However, all of them encounter the same problematic point: the necessity to truncate the series (5). The matter is, the convergence of such kind of series is a very delicate problem, because both signs and relative magnitudes of $A_{i>2}$ coefficients are not predictable *a priori*. Therefore, at finite strains the rule “the more terms the better” may fail, in particular, if the highest order term, survived after truncation, and the successive one (which does not) in fact approximately compensate each other. So, at $Bp$ values close to 4, the third order Birch-Murnaghan EOS (2) may not provide more accurate description, as compared with (7), and, for some cases, even be the worse approximation, see also [5], section 6.4 therein (as a potential “candidate”, in [16] the value of $Bp$=4.001 was theoretically predicted for hypothetical $LiSnO_3$ perovskite). The similar problem can be expected for EOS (10) at small ($Bp$–2) values (so, for $BaVO_3$ the value of $Bp$=2.64 was obtained in [17]).

Another approach to construct EOSs is ground on the particular microscopic models of inter-atomic interactions in solids. The pair inter-atomic potentials, including the interactions of different nature (electrostatic interactions, Born-Meyer repulsion, van der Waals attraction, etc.) are taken into account, and it allows to evaluate directly the total internal energy $U(V)$ of solid (in the simplest case, for a static lattice at zero temperature) – as a background to derive EOS. The simplest example, that we shall consider here, are the noble gases, crystallizing at low temperatures into cubic *fcc* structure. It is well-known that the inter-atomic interactions for them are reliably described with Lennard-Jones potential, $\Phi_{i,j}=4\varepsilon\cdot[(\sigma/r_{ij})^{12}-(\sigma/r_{ij})^{6}]$, where $r_{ij}$ is the distance between i- and j-th atoms, ε and σ are the parameters of potential. For a cubic cell, the distance between two nearest neighbor atoms $R$ can be expressed in terms of volume of unit cell as $R=\gamma\cdot V^{1/3}$, where γ is a geometric factor (for *fcc* lattice $\gamma=2^{-1/2}$), so, it is straightforward to evaluate the total internal energy $U$(V), performing the summation over *fcc* lattice, and to derive the following EOS:

$$P^{(LJ)}(V)=\frac{B_0}{2}\cdot\left[\left(\frac{V_0}{V}\right)^5-\left(\frac{V_0}{V}\right)^3\right] \quad (11)$$

(assuming for simplicity, that the temperature is low enough, and the second term in the expression $F=U-T\cdot S$ can be neglected). The values of $V_0$ and $B_0$ can be expressed here in term of the model parameters as $V_0=[2\cdot(\eta_{12}/\eta_6)\cdot(\sigma/\gamma)^6]^{1/2}$, $B_0=[8\cdot N\cdot\varepsilon\cdot\eta_6\cdot(\sigma/\gamma)^6]/V_0^3$, where N is the quantity of atoms in the unit cell (for *fcc* lattice N=4), and $\eta_6\approx14.45$, $\eta_{12}\approx12.13$ are the constants of summation. The more sophisticated examples have been discussed in [5, 7], in particular, the cases of two- and three-term power law inter-atomic potentials, and of potentials with exponential terms were regarded there. In [18-19] the elastic behavior of some metallic solids and alkaline earth oxides under pressure was theoretically described in terms of model inter-atomic interactions (and some additional contributions to energy of solids). However, this interesting approach requires the comprehensive considerations, which are beyond the topic of the paper, so, we just shall mention its weak side. It is the problem of *transferability* of EOS, obtained in such a manner, to the crystals with other kinds of chemical bonding.

Finally, the empirical way to derive EOSs also takes place, where the regularities in $P$-V behavior, experimentally established and holding more or less reliably for a wide family of compounds, are regarded as an initial suggestion. So, the supposition that the isothermal bulk modulus of solids is proportional to applied hydrostatic pressure, i.e:

$$B = B_0 + Bp \cdot P \,, \qquad (12)$$

is the basic assumption to derive the Murnaghan EOS. In fact, at constant temperature it is possible to put simply, that $B = -\mathrm{V} \cdot (dP/d\mathrm{V})$, further to solve the differential equation (12) at initial condition $P(\mathrm{V}_0) = 0$ and to obtain EOS (1). In the applications, involving *ab initio* modeling of electronic structure of solids, the $U(\mathrm{V})$ form of EOS is usually required. Though in general $(\partial U/\partial \mathrm{V})_\mathrm{T} = [-P + \mathrm{T} \cdot (\partial P/\partial \mathrm{T})_\mathrm{V}]$, at low temperatures the second term can be neglected, and we may simply write out that $P = -dU/d\mathrm{V}$. Integrating (1) at formal initial condition $U(\mathrm{V}_0)=U_0$ (where $U_0$ is a constant, undefined yet) one can obtain:

$$U(\mathrm{V}) = U_0 + \frac{B_0 \cdot \mathrm{V}}{Bp} \cdot \left[ \frac{(\mathrm{V}_0/\mathrm{V})^{Bp}}{Bp-1} + 1 \right] - \frac{B_0 \cdot \mathrm{V}_0}{Bp-1} \,, \quad (13)$$

i.e. the conventional form of $U(\mathrm{V})$ Murnaghan EOS (we also note that though the additive constant of integration may be introduced in various ways, the provided form is recommended for practical fitting of $U(\mathrm{V})$ computational data, because it ensures more robust determination of the fitting parameters, see [20]). For better comprehensiveness, we mention here the other example of the empirical initial proposition for EOS: R. Grover with co-workers [21] have been pointed out that for a wide variety of metals in a wide range of pressures the empirical relation $\ln(B(\mathrm{V})/B_0) = \alpha \cdot (\mathrm{V}_0 - \mathrm{V})/\mathrm{V}_0$ (where $\alpha = Bp$, in fact) takes place. Unfortunately, the EOS $P(\mathrm{V})$ cannot be expressed from that in terms of elementary functions, so, for its practical uses some simplifications are required.

The Murnaghan EOS (1) had been proposed in 1944 and since then is widely employed in geophysics and other fields. Its obvious merits are the relatively simple form and the absence of the necessity to truncate during derivation, making free (at least, formally) from the problem, mentioned above. But how accurate the suggestion (12) is, i.e. is a region of its validity wide enough? At infinitesmall compressions, since both $P \approx -B_0 \cdot \delta$ and $B \approx B_0 + (\partial B/\partial \delta)_{\delta=0} \cdot \delta$ (i.e. depend linearly on $\delta$), it holds exactly. But at finite, not too small deformations? In geophysics, the consistence of EOS with the results, provided by Parametric Earth models (PEM) ([22], see also [7] and Figure 1 therein) is often regarded as a criteria of its plausibility. According to PEM, at high pressures (~50-350 GPa) the bulk modulus of the substance in Earth mantle and core behaves approximately linearly with $P$ (except the discontinuity on the mantle-core boundary and some more peculiarities, irrelevant here), characterizing, though, also by slight curvature, and, consequently, non-zero expected value of $Bpp$ (in PEM it is regarded to be negative). However, within the suggestion (12), of course, $Bpp$=0. It is known, that for a wide variety of materials the Murnaghan EOS reasonably fits experimental data at the pressures up to $\sim B_0/2$ ([5], section 7.4), however, for high compressions it systematically overestimates $P(\mathrm{V})$ values. The widely known approach to improve the EOS (1) is to regard pressure as a small parameter and to expand $B$ formally into series in terms of $P$ , i.e. $B = B_0 + Bp\cdot P + (Bpp/2)\cdot P^2 + \ldots$, typically truncating to the second-order term (see [23] for comprehensive theory). However, what pressures may be regarded as small? For ultra-compressible alkali metals and stiff carbides, they obviously should differ. And, will such truncated expansion be robust at $P > B_0/2$? Alternatively, in [24] the authors have proposed another approach: to construct $(dB/dP)$ dependence in the form $(dB/dP) = \mathrm{n}\cdot[1+\mathrm{m}\cdot f(P)]$, where n, m are the constants, and $f(P)$ are different trial functions of pressure, which tend to zero, if $P$ does. Unfortunately, the EOS, constructed in such a way, becomes highly complicated, as compared with (1), and its advantages in fitting $P$-V data were not shown convincingly there. At the same time, it worth to point out two fine approaches to improve the suggestion (12), which are mentioned in literature sometimes as "*simple modifications of Murnaghan equation*", though the form of resulting EOSs is quite different. The first one results in so-called modified Tait EOS. Though originally its derivation was based on a bit different ideas ([6, 7], see also [25]), it is advisable in the present context to regard the assumption $B = (\mathrm{V}/\mathrm{V}_0)\cdot[B_0 + (Bp+1)\cdot P]$ as the initial concept for it. Since $(\mathrm{V}/\mathrm{V}_0)=(1+\delta)$, the latter also can be interpreted as follows: the bulk modulus consists in two terms, the first one varies linearly with pressure, and the second one is a non-linear additive. The later is negligible at infinitesmall deformations, but becomes significant as compression grows. And after integration we can obtain the equation:

$$P^{(T)}(V) = \frac{B_0}{Bp+1} \cdot \left[ exp\left( (Bp+1) \cdot \left( 1 - V/V_0 \right) \right) - 1 \right], \quad (14)$$

which is known as the modified Tait EOS and involved the same set of the fitting parameters, as EOSs (1)-(4) do. We also mention, that in a series of later papers (see [26-28], for instance) an alternative expression is regarded, which is denoted sometimes as "the Grover-Saxena EOS". However, in fact that expression is just another form of EOS (14) and can be reduced to it. The next approach assumes the more sophisticated dependence on pressure, $B \cdot (\partial B/\partial P)_T^{-1} = (a + b \cdot P)$, and at constant temperature it results in Mao equation [29]:

$$P^{(Mao)}(V) = \left( \frac{a}{b} \right) \cdot \left[ \left( 1 + B_0 \cdot \left( \frac{b-1}{a} \right) \cdot \ln\left( V_0/V \right) \right)^{b/(b-1)} - 1 \right], \quad (15)$$

where $a = B_0/Bp$, and $b = 1 - (B_0 \cdot Bpp/Bp^2)$. The later, as well as EOS (8), contains one more fitting parameter *Bpp*, and if it would be put to zero, we simply would result in Murnaghan EOS. Though the $B(P)$ dependence, more complex than (12), leads to rather complicated form of EOSs, the last two examples show that a supposed relation for bulk modulus not necessarily should have a pressure-only dependent form. It provides a hint for a simple way to improve the EOS (1). But first we shall specify briefly the particular reason, that motivates to do it.

### 2.2 The "flexibility" of EOS

Now we shall consider the problem, mentioned above and common for the most of the regarded EOSs, – the lack of their "flexibility". What does this problem mean? The best way to answer is to illustrate it directly. Fig.1 presents two different parts of computational $U(V)$ dependence for magnesium oxide – the near-equilibrium region (a) and high-pressure region (b) (for computational data, it is more convenient to deal with $U(V)$ dependence, instead of $P(V)$). Both fragments were separately fitted by EOS (13), and the sets of $U_0$, $B_0$, $Bp$, and $V_0$ values were determined for each case. The insets illustrate reproduced behavior of $U(V)$ dependence (13) (with evaluated parameters) in "opposite" region, i.e. in high pressure region for (a) and *vice versa*. As it is seen, though the quality of fitting for each region separately is quite good, the "opposite" regions in both cases are reproduced poorly, moreover, the accordance of the obtained values of fitting parameters (they are also specified in Fig.1) is not satisfactory. The similar situation may be expected also for other regarded EOSs, not only for (13). It is because the EOS is not sufficiently "*flexible*" to reproduce reliably both high-pressure and near-equilibrium regions *simultaneously*. Further, the interesting results of V($P$) measurements up to 20 GPa for $Zn_2SnO_4$ stannate with inverse cubic spinel structure ($F$d-3m), and their fitting by Vinet EOS (3), published by S. Anzellini with co-workers [30], are reproduced in Fig.1(c). The presented fitting, however, includes only the data up to 10 GPa ("which coincides with the limit of quasi-hydrostaticity for the used pressure-transmitting medium", as the authors have pointed out), and the obtained values of parameters are $B_0$=150.0 GPa, $Bp$=7, $V_0$=649.3 Å$^3$. For low-pressure region, the accordance of fitting with experiment is excellent, in other words, EOS (3) at the specified $V_0$, $B_0$, $Bp$ values reproduces the low-pressure behavior of $Zn_2SnO_4$ quite reliably. However, for high-pressure region, the accordance apparently is not satisfactory. So, EOS (3) fails to describe robustly both low- and high-pressure regions with the same set of values of parameters. Achieving the good quality of fitting in low-pressure region, the EOS has no further "*degrees of freedom*" to achieve it at high pressures too. It should be pointed out, however, that the near-equilibrium region has a priority in some sense: First, $V_0$, $B_0$, $Bp$ parameters have the particular physical meaning, and their values characterize the elastic behavior of solids namely in low-pressure region. And next, a lot of the tasks, regarded in the solid state physics, concern the behavior of the solids near the equilibrium state, so, the important demand for EOSs is to describe the near-equilibrium region correctly. The values of $V_0$, $B_0$, $Bp$, obtained from a fitting of a near-equilibrium region only, can be treated as trustworthy quantities, characterizing the elastic behavior of solids near the equilibrium state, but those, obtained from a fitting of a high-pressure region only would be simply fitting parameters with non-obvious physical meaning. For instance, their values, provided in Fig.1(a), are trustworthy in the specified sense, but those, provided in Fig.1(b), are not. The values of $V_0$, $B_0$, $Bp$ for $Zn_2SnO_4$, specified above, are trustworthy, because they allow to reproduce namely the low-pressure region excellently (Fig.1(c)), however, their quantities, obtained from the fitting of high-pressure region only, would be of disputable reliability. The similar situation may be expected also for a fitting of a whole data sampling, including both low- and high-pressure regions, – the formal overall optimization may not ensure the satisfactory accordance in near-equilibrium region. In other words, the "binding" of the EOS to the data, related to the near-equilibrium region, is assumed as a necessary initial step of fitting, otherwise $V_0$, $B_0$, $Bp$ simply may lose their physical meaning. But if the high-pressure region would be reproduced poorly then? It means, the present EOS may require additional "degree of freedom" to do better.

It is time now to make our considerations to be more certain and quantitative. To begin, it is advisable to consider the general formal expansion of pressure and bulk modulus into series in terms of the relative compression δ:

$$P = p_1\cdot\delta + p_2\cdot\delta^2 + p_3\cdot\delta^3 + p_4\cdot\delta^4 + p_5\cdot\delta^5 + p_6\cdot\delta^6 + \ldots,$$

$$B = b_0 + b_1\cdot\delta + b_2\cdot\delta^2 + b_3\cdot\delta^3 + b_4\cdot\delta^4 + b_5\cdot\delta^5 + b_6\cdot\delta^6 + \ldots,$$

where $P_{\delta=0}=0$, $p_k = (\partial^k P/\partial\delta^k)_{\delta=0}/k!$ are the expansion coefficients, and analogously for $b_k$, $\delta = (V/V_0)-1$. At constant temperature, it is easy to obtain that $p_1 = -B_0$, $p_2=(B_0/2)(Bp+1)$, and $p_3 = -(B_0/6)\cdot[(Bp+1)(Bp+2)+B_0\cdot Bpp]$ (taking (9) into account), so:

$$P = -B_0\cdot\delta + (B_0/2)(Bp+1)\cdot\delta^2 - (B_0/6)[(Bp+1)(Bp+2)+B_0\cdot Bpp]\cdot\delta^3 + p_4\cdot\delta^4 + \ldots. \quad (16)$$

Since bulk modulus can be expressed in the form $B = -(1+\delta)\cdot(\partial P/\partial\delta)_T$, it is possible to express $b_k$ coefficients in terms of $p_k$ as $b_k = -[k\cdot p_k+(k+1)\cdot p_{k+1}]$ at $k\geq1$, furthermore, $b_0 = B_0$, $b_1 = -(p_1+2\cdot p_2) = -B_0\cdot Bp$, so:

$$B = B_0 - B_0\cdot Bp\cdot\delta - (B_0\cdot(Bp+1)+3\cdot p_3)\cdot\delta^2 - (3\cdot p_3+4\cdot p_4)\cdot\delta^3 - (4\cdot p_4+5\cdot p_5)\cdot\delta^4 - \ldots \quad (17)$$

(we may use either $p_3$ denotation or the explicit expression for it, where more convenient). The expansions (16), (17) are the most general forms of corresponding series, i.e. free from any model or empirical assumptions, and exhibit the most general regularities, so, we may call them "*right*" (we do not put here the question, how fast do these series converge in such a form, because we further are not going to truncate them). It is very important to note that the terms of lowest orders in (16), (17) depend only on $V_0$, $B_0$, $Bp$ parameters, however, the terms of higher orders, besides them, also include $Bpp$ and $(d^kB/dP^k)_{P=0}$ derivatives of higher orders. In close vicinity of $V=V_0$ the dependence $P(\delta)$ behaves approximately as $-B_0\cdot\delta + (B_0/2)(Bp+1)\cdot\delta^2$, the contributions of terms of higher orders are not significant at rather small δ values. Hence, the values of $V_0$, $B_0$, $Bp$ parameters can be determined independently during the fitting within the specified near-equilibrium vicinity (if it includes sufficient data points for a reliable fitting, though). The values of $V_0$, $B_0$, $Bp$, found in such a way, make EOS able to reproduce the near-equilibrium vicinity accurately, hence, they may be regarded as trustworthy for a whole data region. At their found values (fixed now), the fitting of high-pressure part of the data now can be performed, and the values of other EOS parameters, responsible for higher-orders terms (significant there), can be determined.

Now, it is instructive to compare (16) with the analogous expansions into series of various EOSs, for instance, of (1)–(4):

$$P^{(M)} = -B_0\cdot\delta + (B_0/2)(Bp+1)\cdot\delta^2 - (B_0/6)(Bp+1)(Bp+2)\cdot\delta^3 + \\ + (B_0/24)(Bp+1)(Bp+2)(Bp+3)\cdot\delta^4 + ...,$$

$$P^{(BM3)} = -B_0\cdot\delta + (B_0/2)(Bp+1)\cdot\delta^2 - (5\cdot B_0/54)(18\cdot Bp-25)\cdot\delta^3 + \\ + (25\cdot B_0/216)(31\cdot Bp-68)\cdot\delta^4 + ...,$$

$$P^{(VR)} = -B_0\cdot\delta + (B_0/2)(Bp+1)\cdot\delta^2 - (B_0/216)(27\cdot Bp^2+90\cdot Bp+91)\cdot\delta^3 + \\ + (B_0/432)(9\cdot Bp^3+63\cdot Bp^2+167\cdot Bp+161)\cdot\delta^4 + ..., \quad (18)$$

and

$$P^{(AS)} = -B_0\cdot\delta + (B_0/2)(1-2\cdot n)\cdot\delta^2 - B_0((n^2/2)-n+1/3)\cdot\delta^3 + \\ + B_0((3\cdot n^2/4)-(11\cdot n/12)-(n^3/6)+1/4)\cdot\delta^4 + ...$$

(where $n = -Bp/2$), respectively. It is seen that first two terms are the same everywhere and coincide with those of general expansion (16), this way, they are "*right*" and require no modification, this is the general property for all physically consistent EOSs, applicable to solids. The discrepancy begins from the third term, proportional to $\delta^3$, and further. For each EOS, the parameters $V_0$, $B_0$, $Bp$, characterizing the near-equilibrium properties, predefine in its own particular way the higher order terms, because their various combinations actually enter there, instead of *Bpp* and higher order parameters. In other words, the low-pressure behavior of EOSs predefines their behavior at high pressures. And it is the principal reason of the mentioned lack of "flexibility" of EOSs. Any of them can be good to describe the near-equilibrium elastic behavior of a material (as $\delta$ is small), but then there remains no "degree of freedom" to reproduce reliably its high pressure behavior. Some proposed variants of such replacements of coefficients in higher order terms, of course, may provide more or less suitable approximation even for wide class of materials, however, it is just happy, but random coincidence. Most likely, there is no universal and "*uniquely right*" relation between $p_k$ coefficients at higher order terms in (16), and each of them, in general, may be regarded as an additional potential "degree of freedom" making EOS more "flexible". For comparison, we shall also provide here the expansion of Mao EOS (15), involving one more fitting parameter:

$$\begin{aligned} P^{(\mathrm{Mao})} = & - B_0 \cdot \delta + (B_0/2)(Bp+1)\cdot\delta^2 - (B_0/6)(Bp^2 + 3\cdot Bp + 2 + B_0\cdot Bpp)\cdot\delta^3 + \\ & + (B_0/24)(Bp^3 + 6\cdot Bp^2 + 11\cdot Bp + 6 + 3\cdot(2+Bp)\cdot B_0\cdot Bpp + (2/Bp)(B_0\cdot Bpp)^2)\cdot\delta^4 + \dots, \end{aligned} \tag{19}$$

(where the parameters *Bp*, *Bpp* are involved instead of a and b, see above). The *Bpp* parameter appears in third and higher order terms. Its presence provides a little effect in near equilibrium region, but becomes important at not too small $\delta$. The possibility to vary its value allows to correct the contributions of higher order terms and improves the high-pressure "flexibility" of EOS. However, the fourth and higher order terms in expansion (19) are still predefined – though, by four parameters $V_0$, $B_0$, *Bp* and *Bpp* now. If such additional "degree of freedom", the EOS possesses now, would appear to be insufficient, our consideration can be expanded in obvious way, and additional fitting parameters, responsible for higher orders terms, may be introduced.

Summarizing, it seems advisable to perform practically the fitting in two-steps process: first to determine the parameters $B_0$, *Bp*, $V_0$ from the fitting of near-equilibrium region, then to expand the data sampling, adding the high-pressure region, and, using the results of the first step, to determine the optimal values of other parameters, responsible for high-pressure behavior. Further, the values of parameters can be refined in iterative way.

# 3 The improved Murnaghan EOS

Keeping in mind the regards above we shall consider the certain way to make the Murnaghan EOS more precise at high pressures. Let us return to the initial suggestion (12), taking the expansion (16) into account, it can be rewritten as:

$$B = B_0 - (B_0 \cdot Bp) \cdot \delta + (B_0 \cdot Bp(Bp+1)/2) \cdot \delta^2 + Bp \cdot p_3 \cdot \delta^3 + Bp \cdot p_4 \cdot \delta^4 + \ldots \quad (20)$$

Comparing this expression with the general expansion for bulk modulus (17) we see that their first two terms coincide, i.e., they are "*right*", the "dissention" starts from the second order terms (unless $p_3 = -B_0 \cdot (Bp+1) \cdot (Bp+2)/6$) and further. However, the second order term of expansion (20) can be "*corrected*", for this purpose we shall suggest, instead of (12), that:

$$\begin{aligned} B &= B_0 + Bp \cdot P - [(B_0 Bp(Bp+1)/2) \cdot \delta^2] + [-(B_0(Bp+1) + 3 \cdot p_3) \cdot \delta^2] = \\ &= B_0 + Bp \cdot P - [B_0(Bp+1)(Bp+2)/2 + 3 \cdot p_3] \cdot \delta^2 \end{aligned} \quad (21)$$

i.e. subtract its "*incorrect*" version and add the "*right*" one. The expansion of (21) into series will coincide with (17) already up to second order terms. Such kind of modification "*embeds*" the new quantity into the expression for bulk modulus, – $p_3$, which is regarded as independent on $B_0$, $Bp$, $V_0$, i.e. as an additional independent fitting parameter. Of course, not only the second order term in (20) can be "*corrected*" in this way, the higher order terms can be too, i.e.:

$$\begin{aligned} B &= B_0 + Bp \cdot P - [(B_0/2)(Bp+1)(Bp+2) + 3p_3] \cdot \delta^2 - [(Bp+3)p_3 + 4p_4] \cdot \delta^3 - \\ &- [(Bp+4)p_4 + 5p_5] \cdot \delta^4 - \ldots - [(Bp+\mathrm{N})p_\mathrm{N} + (\mathrm{N}+1)p_{\mathrm{N}+1}] \cdot \delta^\mathrm{N} \end{aligned} \quad (22)$$

where as many terms of the series (20) are modified as advisable, but a finite number (because modification of a finite number of terms does not affect the convergence of a series qualitatively, just changes its sum). Writing again the bulk modulus in the form $B = -(1+\delta) \cdot (dP(\delta)/d\delta)$, at constant temperature, and taking the expression (22) into account, it is straightforward to obtain the following differential equation:

$$\frac{dP(\delta)}{d\delta} + Bp \cdot \frac{P(\delta)}{1+\delta} = f_{\mathrm{N}}(\delta), \quad (23)$$

where

(24)

$$f_{\mathrm{N}}(\delta) = -\frac{B_0}{(1+\delta)} + \left[B_0(Bp+1)(Bp+2)/2 + 3p_3\right] \cdot \frac{\delta^2}{(1+\delta)} + \sum_{\mathrm{i}=3}^{\mathrm{N}} \left[(Bp+\mathrm{i})p_{\mathrm{i}} + (\mathrm{i}+1)p_{\mathrm{i}+1}\right] \cdot \frac{\delta^{\mathrm{i}}}{(1+\delta)}$$

It can be solved with the variation of constants method at the initial condition $P(\mathrm{V}_0)=0$, its general solution can be found in the form $P(\delta) = \mathrm{C}(\delta)\cdot(1+\delta)^{-Bp}$, where $\mathrm{C}(\delta)$ is a function, unknown yet. The further details of the solving, as well as some qualitative remarks, related to the development of EOS, can be found in the Supplementary materials (SupM). Here we just provide and discuss the solutions for N=2 and 3, as the most practically important result.

Taking N=2 and introducing for further the dimensionless parameter α as follows:

$$\alpha = 1 + \frac{6 \cdot p_3}{B_0 \cdot (Bp+1)(Bp+2)}, \quad (25)$$

we can write out the solution of (23) in the form:

$$P(\delta) = \frac{B_0}{Bp} \cdot (1-\alpha) \cdot \left[\frac{1}{(1+\delta)^{Bp}} - 1\right] + \alpha \cdot B_0 \cdot \left[\frac{(Bp+1)}{2} \cdot \delta^2 - \delta\right],$$

or, in terms of $(\mathrm{V}/\mathrm{V}_0)$ ratio:

$$P(\mathrm{V}) = \frac{B_0}{Bp} \cdot (1-\alpha) \cdot \left[\left(\frac{\mathrm{V}_0}{\mathrm{V}}\right)^{Bp} - 1\right] + \alpha \cdot B_0 \cdot \left[\frac{(Bp+1)}{2} \cdot \left(\frac{\mathrm{V}}{\mathrm{V}_0} - 1\right)^2 - \left(\frac{\mathrm{V}}{\mathrm{V}_0} - 1\right)\right]. \quad (26)$$

The equation (26) is similar to the Murnaghan EOS (1) and, obviously, turns into it, if α=0. In that case the relation $p_3 = -(B_0/6)(Bp+1)(Bp+2)$ would occur, as it takes place in the expansion (18) of Murnaghan EOS, and the "*correction*" to bulk modulus in (21) then would equal to zero. This way, the EOS (26) may be interpreted as follows: it results from the Murnaghan EOS, multiplied by the correction factor (1–α), meanwhile first two terms in its expansion are protected from such modification (due to appearance of the "tail" $\alpha\cdot B_0\cdot[\ldots]$), because, as discussed above, they are "*right*" and require no corrections. The discrepancy between (26) and (1) becomes significant, as $\delta$ and, respectively, higher order terms of $P(\delta)$ series grow. The variations of $\alpha$ parameter modify the coefficients at higher order terms in $P(\delta)$ expansion, providing the additional "flexibility" of (26) in the fitting, as compared with (1). Generally, there seems to be no principal limitations both for sign of $p_3$ and value of $p_3/B_0$ ratio (at least, following from general considerations), so, the same can be expected for $\alpha$ parameter. For potential comparison with other EOSs, it is convenient to write out here the expression for *Bpp*, obtained from (26), as:

$$Bpp = -\alpha\cdot(Bp^2 + 3\cdot Bp + 2) / B_0,$$

We note that for geophysical applications (and in PEM, in particular) [7] the condition $Bpp < 0$ is regarded as the criteria of plausibility of EOSs (assuming that $Bp > 0$), so, the requirement $\alpha > 0$ would follow from there. At the same time, for crystalline solids, the situations beyond the mentioned "geophysical" criteria also may take place, and we shall keep it in mind. Finally, the EOS (26) is not too complicated, as compared with (1), at least, it is simpler in form than fourth-order Birch-Murnaghan EOS (8) or Mao EOS (15) (Besides, the $U$(V) form of the latter cannot be expressed in terms of elementary functions. It is reduced to the integral, analogous to that determining Γ-function, and this complicates its practical employ).

For N=3, the solution of (23) results in the third-order EOS as follows:

$$P(\mathrm{V}) = \frac{B_0}{Bp}\cdot(1-\alpha)\cdot\left[\left(\frac{\mathrm{V}_0}{\mathrm{V}}\right)^{Bp} - 1\right] + \alpha\cdot B_0\cdot\left[\frac{(Bp+1)}{2}\cdot\left(\frac{\mathrm{V}}{\mathrm{V}_0}-1\right)^2 - \left(\frac{\mathrm{V}}{\mathrm{V}_0}-1\right)\right] + \lambda\cdot B_0\cdot\left(\frac{\mathrm{V}}{\mathrm{V}_0}-1\right)^3, \quad (27)$$

where

$$\alpha = 1 - \frac{24\cdot p_4}{B_0\cdot(Bp+1)(Bp+2)(Bp+3)}, \qquad \lambda = \frac{1}{B_0}\cdot\left[p_3 + \frac{4\cdot p_4}{(Bp+3)}\right]$$

are the newly introduced dimensionless parameters. Obviously, if $\lambda=0$, thereat $4\cdot p_4 = -(Bp+3)\cdot p_3$, and EOS (27) turns into second-order EOS (26), it would correspond to zero value of the correction, proportional to $\delta^3$, in (22). It is seen from the structure of EOS (27) that its base idea was inherited from (26), however, in addition, the coefficient at third-order term in $P(\delta)$ expansion can be fitted individually due to variations of $\lambda$ parameter. In this manner, the order of EOS can be further increased, if advisable. It will be shown further, though, that even second-order EOS will be sufficient to describe quite satisfactorily the $U$(V) or $P$(V) behavior of crystalline solids, regarded as examples.

# 4 The application to the computational and experimental data

## 4.1 The two-steps fitting process

In the present section we shall regard the application of proposed modified Murnaghan EOS to fit the data, obtained in *ab initio* modeling or in experiment. For the treatment of the experimental $P$(V) dependence the EOS (26) can be applied directly. And for the fitting of computational data, obtained within the simple suggestion of static crystal lattice, it should be converted first to $U$(V) form – in the same manner, as (13) had been derived by integration of (1), i.e.:

$$
\begin{aligned}
U(\mathrm{V}) = U_0 + \frac{B_0 \cdot \mathrm{V}}{Bp} \cdot (1-\alpha) \cdot \left[ \frac{(\mathrm{V}_0/\mathrm{V})^{Bp}}{Bp-1} + 1 \right] - \frac{B_0 \cdot \mathrm{V}_0 \cdot (1-\alpha)}{Bp-1} - \\
- \alpha \cdot B_0 \cdot \mathrm{V}_0 \cdot \left[ \frac{(Bp+1)}{6} \cdot \left( \frac{\mathrm{V}}{\mathrm{V}_0} - 1 \right)^3 - \frac{1}{2} \cdot \left( \frac{\mathrm{V}}{\mathrm{V}_0} - 1 \right)^2 \right]
\end{aligned}
\quad (28).
$$

As the examples of experimental data, suitable for the fitting, the results of very detailed V–$P$ measurements for metallic chromium [31] and ruthenium [32], as well as the mentioned above data for $Zn_2SnO_4$ stannate [30], were taken. To test the applicability of (28) to computational $U$(V) data, the following crystalline solids with cubic structure were chosen as a representative selection: α-calcium (space group *Fm-3m*), magnesium oxide MgO (*Fm-3m*), titanium nitride (*Fm-3m*), crystalline diamond and silicon (*Fd3m*) and $SrTiO_3$ oxide perovskite (*Pm-3m*). All the calculations of their electronic structure have been performed within the full-potential linearized method of augmented plane wave (FP-LAPW) [33], where the effects of exchange and electronic correlation were taken into account within the generalized gradient approximation (GGA) in Perdew-Burke-Ernzerhof (PBE) form [34]. The software package Wien2k [35], where the mentioned computational method had been implemented, was used for the calculations. For the selected compounds, no especial computational schemes are required to be involved, the calculations of their electronic structure are rather ordinary, and their description can be found elsewhere in manifold papers, so, we shall omit them (see, though, the SupM for some details, as well as for the obtained numerical $U$(V) data). Note, during the modeling we ignore any possible structural phase transitions, that could occur at elevated pressures, because this is a separate topic, related to the present one indirectly.

As it was mentioned before, the two-steps fitting process (with "binding" to near-equilibrium region) can be advisable here. At the first stage, for each considered compound, we shall select in the attributed $\{V_i, P_i\}$ or $\{V_i, U_i\}$ sampling its sub-sampling – the near-equilibrium region. For $\{V_i, U_i\}$ it should represent the vicinity of the minimum of $U$(V) dependence, for $\{V_i, P_i\}$ – the segment with upper boundary at $V_i$ value, corresponding to $P_i$, nearest to zero. The chosen region of data will be fitted with the ordinary Murnaghan EOS (1) or (13), respectively (or by (26), (28) at $\alpha$=0, it is the same) – to obtain the "preliminary" values of $U_0$, $B_0$, $Bp$, $V_0$ parameters therefrom. Here, the reasonable question on the quantitative criteria for such choice of near-equilibrium region arises. The principal demand, underlying the answer, is: the Murnaghan EOS (13) should behave within selected region still as $-B_0\cdot(V/V_0-1) + (B_0/2)(Bp+1)(V/V_0-1)^2$ with acceptable accuracy, i.e. the higher order terms should be still insignificant therein. The boundary should separate it from the region, where it becomes advisable to take the higher order terms into account. It can be formulated quantitatively in various forms, however, the simplest one can be proposed as follows. 1) Within the selected trial region the condition

$$\left| \frac{P^{(1)}(V)}{-B_0(V/V_0-1)+(B_0/2)(Bp+1)(V/V_0-1)^2} - 1 \right| \le \varepsilon$$

should be satisfied, where $\varepsilon$ defines the acceptable limit of deviation of $P^{(1)}$(V) dependence (1) from its truncated approximation within the near-equilibrium region. The fitting of a trial sub-sampling provides the trial values of $B_0$, $Bp$, $V_0$ – to check, if the proposed criteria is satisfied. And 2) The selected trial region should cover enough data points to perform a robust fitting, i.e. the obtained trial values of $B_0$, $Bp$, $V_0$ should not be too sensitive to small variations of its boundaries. Generally, the values ε ~ 0.03–0.05 can be recommended, however, they may also depend on properties of a specific sampling. Note also, though for particular solids, the Murnaghan EOS may provide the good agreement in region, considerably wider than the specified one, we proceed from the general considerations and outline the demands, as general as possible. Though they provide no strict boundaries, their reasonable estimations are quite possible. Some more aspects, concerning the selection of such near-equilibrium region, are regarded in SupM for some particular examples.

At the next stage of fitting we shall take the whole sampling and, accepting the "preliminary" $B_0$, $Bp$, $V_0$ values, obtained at the previous stage (for $\{V_i, U_i\}$, $U_0$ parameter also is added), find the optimal value of α parameter, providing the best fit. It is straightforward to derive the following formula:

$$\alpha = \frac{\sum_{i=1}^{N}[U^{(13)}(V_i) - U_i]\cdot\left[U^{(13)}(V_i) - U_0 + \phi(V_i)\right]}{\sum_{i=1}^{N}\left[U^{(13)}(V_i) - U_0 + \phi(V_i)\right]^2},$$

$$\phi(V_i) = B_0 V_0 \cdot \left[\frac{(Bp+1)}{6}\cdot(V_i/V_0 - 1)^3 - \frac{1}{2}\cdot(V_i/V_0 - 1)^2\right], \qquad \text{and}$$

$$\alpha = \frac{\sum_{i=1}^{N}[P^{(1)}(V_i) - P_i]\cdot\left[P^{(1)}(V_i) - \psi(V_i)\right]}{\sum_{i=1}^{N}\left[P^{(1)}(V_i) - \psi(V_i)\right]^2}, \tag{29}$$

$$\psi(V_i) = B_0 \cdot \left[\frac{(Bp+1)}{2}\cdot(V_i/V_0 - 1)^2 - (V_i/V_0 - 1)\right],$$

for the fitting of $\{V_i, U_i\}$ and $\{V_i, P_i\}$ data, respectively, where $P^{(1)}$ and $U^{(13)}$ mean the expressions (1) and (13), i =1,2…N, i.e. goes over all the points of the sampling. Further, the obtained values of the parameters can be refined in iterative way, if necessary, as follows: the near-equilibrium region of the sampling (not the whole one!) is fitted by (26) or (28) at constant, previously determined α value, – to precise the values of $B_0$, $Bp$, $V_0$ parameters (and $U_0$, where necessary), then, with their refined values, the value of $\alpha$ is improved by formula (29), and so on, this cycle is repeated till the desired accuracy will be achieved. We note again that the attempt to fit at once the whole sampling by (26) or (28) in a formal manner highly risks to result in unreliable values of parameters. To avoid it, the described two-steps fitting process was proposed.

### 4.2 The computational *U*(V) data

The behavior of *U*(V) dependence, calculated for each of the representative compounds, i.e. for $\alpha$-Ca, MgO, TiN, diamond, silicon and $SrTiO_3$, is visualized in Fig.2, both near-equilibrium region (the vicinity of the minimum) and high-pressure region are included there. The near-equilibrium regions, selected in the specified above way, are marked by green color both in Fig.2 and in the Tables in SupM. Furthermore, besides the fitting of the full $\{V_i, U_i\}$ sampling, we also shall attempt to explore "the predictive possibilities" of the improved Murnaghan EOS (28). It means the following: we have selected the near-equilibrium region previously, now we shall select also the adjacent region, which corresponds to intermediate, not too high pressures (marked by blue color both in Fig.2 and in the Tables in SupM). The "*blue* + *green*" sub-sampling will be treated as "shortened", and the described two-steps fitting procedure will be also applied to it – to find out, how reliably EOS (28) with parameters, obtained in such a way, can predict the behavior of *U*(V) beyond the shortened sampling (as compared with the result for the full sampling).

The comparative results of fitting by ordinary (13) and improved Murnaghan EOS (28) both of full and shortened $\{V_i, U_i\}$ sampling also are visualized in Fig.2 (magenta, black and orange lines, respectively), and the obtained fitting parameters $B_0$, $Bp$, $V_0$ and $\alpha$ are presented in the Table 1. For each regarded compound, it is seen on the insets, that both EOS (13) and (28) excellently fit the "*green*" near-equilibrium region, i.e. the condition of "binding" of EOS to the near-equilibrium region is satisfied. The noticeable deviation of ordinary Murnaghan *U*(V) EOS (13) from the computed $\{V_i, U_i\}$ sampling starts at some moderate pressure (whose approximate values are specified in Fig.2), increasing with its further growth. Comparing with equilibrium bulk modulus, it can be seen, that for various compounds with different kinds of chemical bonding the values of "deviation pressure", specified in Fig.2, may observably differ from the mentioned-above evaluation ~$B_0/2$. So, for MgO the reasonable accordance of EOS (13) with the computed data takes place up to pressures, rather close to $B_0$, but for TiN and α-Ca these estimated pressures are about three times less. It is obvious from Fig.2 that for all the considered compounds the improved $U$(V) dependence (28) fits the computed {$V_i$, $U_i$} sampling much more accurately than (13) – up to high pressures, which are at least not less than $B_0$, or even ~1.2–2 times greater. For MgO and TiN (28) reproduces the computational data reliably up to pressure values ~550–600 GPa, i.e. even beyond those practically reasonable in geophysics. The mentioned advantage is so evident, that it even would be excess to compare the root mean square deviations (RMSD) for (13) and (28). Though such result is expectable, because we started from EOS (13) and *improved* it, providing the additional "flexibility" at high pressures (due to additional parameter α) and keeping the successful fitting in near-equilibrium region at the same time. Moreover, for the most of the regarded compounds, EOS (28) with the values of parameters, optimized for the shortened "*blue + green*" sub-sampling (orange line in Fig.2, compare also the corresponding columns in Table 1), predicts the further $U$(V) behavior beyond it quite reliably (at least, much better, than EOS (13) does). So, our primarily goal is achieved – as seen in Fig.2, the proposed simple improvement of Murnaghan EOS allows to broaden considerably a region of its applicability.

However, for α-Ca the high-pressure region is reproduced by EOS (28) not so accurately, as compared with other compounds (though it is still much better than the fitting by ordinary EOS (13) is), and its "predictive possibility" appears to be rather poor here. Moreover, beyond the minimum, i.e. at $V>V_0$, (28) reproduces $U$(V) behavior for α-Ca even worse than (13) does (as well as for TiN). Though this region, corresponding to "negative hydrostatic pressures", is of little practical interest and may be "sacrificed", this point hints that for some substances even the "*flexibility*" of (28) may be insufficient to reproduce $U$(V) behavior in a whole region flawlessly. As an attempt to solve the outlined problem, let us regard the fitting of $\{V_i, U_i\}$ data for α-Ca by improved Murnaghan EOS, including two fitting parameters α and λ, responsible for higher-order terms, – in comparison with (28). Its $U(V)$ form may be represented as follows:

$$U(\mathrm{V}) = U^{(28)}(\mathrm{V}) - \lambda \cdot \frac{B_0 \cdot \mathrm{V}_0}{4} \cdot \left(\frac{\mathrm{V}}{\mathrm{V}_0} - 1\right)^4, \quad (30)$$

and some details of the fitting procedure can be found in SupM. The optimized values of its parameters are: $B_0$=18.3392 GPa, $Bp$=4.6405, $V_0$=286.6142 bohr$^3$, α=0.6592, λ=3.8508, and the results of the fitting are visualized in Fig.3 (the black and magenta lines correspond to EOS (28) and (30), respectively). Unfortunately, the increase of the order of EOS does not improve its accordance with $\{V_i, U_i\}$ data at $V>V_0$, moreover, in that region (30) reproduces them even worse than (28) does, though at positive hydrostatic pressure (i.e. at $V<V_0$) the third order EOS yet provides some quantitative fitting advantages (the values of RMSD, calculated for EOS (28) and (30) within the region of $V<V_0$, are $11.75\cdot10^{-4}$ Ry and $6.04\cdot10^{-4}$ Ry, respectively). The reasons of mentioned deviation of EOS (28) from $\{V_i, U_i\}$ data at $V>V_0$ will be discussed in the next section.

At last, it seems to be useful, for comprehensive considerations, to compare the values of bulk modulus $B_0$ and its pressure derivative $Bp$, obtained from the fitting of $U(V)$ computational data, with known experiment. For α-Ca, the results of V–$P$ measurements up to 19.8 GPa (above it turns into *bcc* β-phase) are available [36, 37], and the performed fitting results in the values (at ambient temperature) $B_0$=17.4 GPa, $Bp$=3.22, $V_0$=176.370 Å$^3$ and $B_0$=16.0–16.4 GPa, $Bp$=3.20–3.32, $V_0$=175.4–175.6 Å$^3$ (at various fitting schemes), respectively. Generally, the value 5.5884 Å is specified for a lattice parameter of *fcc* α-Ca in reference books (see [38], for instance), so, its experimental room-temperature value of $V_0$ is expected to be lower, 174.53 Å$^3$. For close-packed structures, it is reasonable to expect the density of single-crystal (assumed within *ab initio* modeling ) to be higher than that of samples in other forms. On the other hand, the *ab initio* structural optimizations, performed at static lattice, ignore the effect of thermal expansion. For metallic calcium, the linear thermal expansion κ is 22.3 (in μm/(m·K) units) at ambient conditions [39], it is noticeably higher, than for iron (~10-12), chromium (~6), silicon (~2.5), diamond (~1.3), TiN (~9.3), $SrTiO_3$ (~9.4), for instance. Taking the computational $V_0$ value 169.72 Å$^3$ (Table 1) as its magnitude at 0 K and "extrapolating" it to 300 K, we obtain $V_0^{(300\ K)} \approx V_0^{(0\ K)} \cdot (1+3\kappa T) \approx 173.13$ Å$^3$. Of course, such estimation is very rough, because κ is a temperature-dependent parameter. However, it shows, that the reported difference between the value of $V_0$, obtained from the fitting of V–*P* measurements, and the calculated one, is reasonable and can be explained. The difference in the values of $B_0$ and *Bp* also can be attributed to thermal effects – typically, the bulk modulus decreases, as temperature grows, in [39] its room-temperature value was reported to be 17 GPa. On the contrary, for other compounds from our representative selection, the calculations overestimate the values of $V_0$ and underestimate those of $B_0$ for them, as compared with available experiment. For diamond, though, taking widely known value 3.56–3.57 Å for its lattice constant, we obtain a good agreement with the calculated $V_0$ value: 45.56 Å$^3$. For silicon, the lattice constant is 5.43 Å, and the obtained theoretical $V_0$ value 163.73 Å$^3$ overestimates the actual one. For bulk TiN, the calculated value 76.63 Å$^3$ agrees reasonably with lattice parameter values 4.23–4.24 Å, typically reported. For MgO, the available experimental values of $V_0$ at room-temperature are in the range 74.69–74.73 Å$^3$ [40-42], for $SrTiO_3$, it was measured to be 59.50 Å$^3$ [43], these quantities are lower than the corresponding theoretical $V_0$ values (Table 1). Generally, the LDA/GGA-based *ab initio* calculations tend to overestimate the optimized structural parameters, as compared with experiment (despite the thermal expansion, if it is moderate). As concerns the elastic characteristics, the paper [44] reports the rather wide spread of their experimental values in the ranges $B_0$=159.6–168.3 GPa, *Bp*=2.5–4.44 for MgO, and $B_0$=82.0–100.08 GPa, *Bp*=1.7–15.3 for silicon (see the References therein). The values of $B_0$ for MgO were determined from *P*–V–T measurements in [40-42] to be in the range 159.0-164.1 GPa, exceeding the value 147.60 GPa, predicted from the calculations, though, the values of *Bp* were found in the range 3.74–4.37, and the theoretical value 4.14 falls within it. For silicon and diamond, the measurements of velocities of acoustic waves, providing the accurate values of their $C_{11}$, $C_{12}$, $C_{44}$ elastic constants (as well as their temperature dependence), are available, [45] and [46], respectively. The value of $B_0$ for diamond was obtained therefrom to be 442.33 GPa, that agrees reasonably with the theoretical result: 440.96 GPa (the close value 442.66 GPa was also obtained in [47] from the experiments on Brillouin scattering). For silicon, the reported $B_0$ value 97.88 GPa noticeably exceeds the theoretically predicted one (88.75 GPa), however, the obtained values of bulk modulus derivatives *Bp* are found to be in surprisingly good agreement, 4.24 [45] *versus* 4.22 (Table 1). One more paper [48], based on the analysis of known experimental data, concerning third-order elastic constants $C_{ijk}$, also provides a close value of *Bp* for silicon to be 4.19. As for TiN, the results of V–*P* measurements for its nanocrystalline samples with different size of the particles (16-80 nm), as well as for powders (with grain sizes ~5 μm) were presented in [49] and [50], respectively. The fitted value of $B_0$ for nanocrystalline TiN varies in the range 287–320 GPa, being minimal at highest particles size (80 nm), and this value is relatively close both to that, reported for powder TiN form (282 GPa), and to the value, obtained from *ab initio* calculations (281.55 GPa). The latter aspect allows to suggest this quantity to be close to the value of $B_0$ for bulk titanium nitride. Unfortunately, the fitting of V–*P* data in both investigations was performed at fixed value of *Bp*=4, though [50] suggests, that the bulk modulus behaves approximately as $B = 275.4 + 3.5 \cdot P$ (in GPa) in the pressure range ~6–45 GPa. The value of the equilibrium bulk modulus for strontium titanate, obtained within the resonant ultrasound spectroscopy experiments (174.9 GPa), is provided in [51]. The results of measurements of velocities of ultrasound waves in $SrTiO_3$, providing the values of $C_{11}$, $C_{12}$, $C_{44}$ elastic constants, are presented in [52], wherefrom $B_0$ can be determined as $(C_{11}+2 \cdot C_{12})/3$ to be 174.03 GPa. These values of $B_0$ are reasonably close to the computational one (170.07 GPa), however, they are related to ambient temperature. The results of [52] allow also to estimate the value of $B_0$ at 110 K (i.e. near the *Pm-3m* → *I4/mcm* structural phase transition at ~106-108 K) to be 179.42 GPa. In other words, the LDA/GGA computations still underestimate $B_0$ with respect to experiment. Though, the detailed comparisons require the separate consideration.

### 4.3 The experimental *P*(V) data

At last, we shall consider the application of the modified Murnaghan EOS (26) to approximate the available experimental data for some representative solids, mentioned above. Fig.4 presents the data of V–*P* measurements for metallic chromium [31] and ruthenium [32], performed in the wide range of pressures – from ambient value to ~150 GPa, as well as the results of their fitting with the proposed EOS (26) in comparison with Birch-Murnaghan EOS (2) of third order (BM3) and Vinet EOS (3). It is seen that all three EOSs successfully fit the *P*–V data in the whole range with no considerable deviations of separate points or data regions, and all three curves visually practically coincide. The values of fitting parameters, obtained in present work for EOS (26), and those, reported for Vinet and BM3, are provided in Table 2. It is seen that for chromium EOS (26) results in $B_0$ value, a bit higher (by 7-10 GPa), and lowers *Bp* value, as compared with other EOSs, for ruthenium the inverse situation takes place, though this difference is within the reasonable range. The values of $\alpha$ parameter of EOS (26), responsible for high-pressure behavior, were obtained to be comparable for chromium and ruthenium, 0.204 and 0.243, respectively. For chromium data, the calculated values of RMSD vary in the range ~0.49-0.56 GPa for all three EOS, for EOS(26) its value is even a bit lower, than for Vinet. At the same time, for ruthenium data, the RMSD for EOS (26) appears to be higher, than for others (~0.69 GPa *vs* 0.47 GPa). Though, the *P*–V measurements for ruthenium are characterized with considerably higher uncertainty in pressure (from 0.05 GPa at low pressures to 2 GPa at *P*~150 GPa [32]), as compared with those for chromium (from 0.06 GPa at ambient pressure to 0.27 GPa at P~200 GPa [31]), and such difference in RMSD can be treated to be within the experimental uncertainty range. This way, EOS (26) reproduces the proposed experimental V-*P* data with no explicit drawback, at least, not worse, than Vinet or BM3 do. On the other hand, it can be seen from the presented results, that even EOS of third order, including only $B_0$, *Bp*, $V_0$ parameters, will be quite sufficient to reproduce accurately the regarded experimental data, anyway, EOSs (2) and (3) do it robustly at pressures up to 150 GPa. The following explanation can be proposed: it is known, that for a lot of metals the empirical relation of R. Grover [21], mentioned above, takes place in a wide range of pressures, namely, $\ln(B(V)/B_0) = Bp\cdot(V_0-V)/V_0$. Though the corresponding $P$(V) EOS cannot be represented in terms of elementary functions, it would not be a problem to write out its expansion:

$$P(\delta) = -B_0\cdot\delta + (B_0/2)(Bp+1)\cdot\delta^2 - (B_0/6)(Bp^2+2\cdot Bp+2)\cdot\delta^3 + \dots$$

Comparing with those for Vinet and BM3 EOSs (18), it can be found, that at $Bp$ values from reasonable range (~3–10) the terms $\sim\delta^3$ are rather close, the ratio of their values is within ~1–1.4. Though EOSs (2) and (3) may occur to be not sufficiently “flexible”, the predefined form of their higher-order terms appears to be quite suitable to provide the accurate approximation of experimental V–$P$ data for chromium, ruthenium, and for a lot of other metals, as it is seen from [53] (V–$P$ data for Co, Ni, Zn, Mo, Ag up to 100-150 GPa were successfully fitted with Vinet EOS) and [54] (K, Al, Be, Au, Cu, Co, Ta, Mo, W, with BM3). Of course, the improved Murnaghan EOS (26) successfully fits the proposed data too, but its enhanced “flexibility” may not to be the necessary demand in similar cases (see also [55] and Fig.6 therein).

However, the example, where its “flexibility” is required definitely, has already been regarded above. It means the V–$P$ measurements for $Zn_2SnO_4$ stannate [30], where the Vinet EOS (3) was unable to fit them successfully in a whole range (Fig.1(c)), even at comparatively moderate pressure, up to 20 GPa. In Fig.5 the results of fitting of the V-$P$ data for $Zn_2SnO_4$ with EOS (26) are presented in comparison with Birch-Murnaghan EOS (8) of fourth order (BM4) and Mao EOS (15), the obtained values of the fitting parameters are collected in Table 2. It is seen that up to ~ 7 GPa all the regarded EOSs behave almost identically to Vinet EOS (which also was plotted for obviousness). Further, up to ~10 GPa, the Vinet EOS still fits the experimental points accurately, Mao EOS also goodly reproduces them in that region, but both BM4 and EOS (26) do it worse, deviating there. Above 10 GPa the Vinet EOS deviates unacceptably, Mao EOS also provides poor quality of fitting there, but BM4 and EOS (26) correspond to the experimental data much better. The calculated overall RMSD value for Mao EOS is 0.446 GPa, and for BM4 and EOS(26) they are 0.204 GPa and 0.227 GPa, respectively. Thus, according to the present example, the fitting abilities of the latter two are comparable, and, in general, they considerably exceed that of Mao EOS (though Mao EOS may "win" locally, in the region 7-10 GPa at present case). The quality of fitting can be slightly improved, if we would employ EOS (27) including the parameters $\alpha$ and $\lambda$ (see previous sub-section for analogy), instead of (26) (the "wine color" plot in Fig.5), the corresponding RMSD is 0.197 GPa. It does better than EOS (26) and even BM4, however, such improvement is not decisive, the region 7-10 GPa still is a weak point in present example, for instance. Nevertheless, it was shown that EOS (26), as well as BM4, is rather "flexible" to reproduce robustly the "difficult" data sampling as a whole (where Vinet EOS and other equations of third order are unable), keeping simultaneously the plausible behavior in the near-equilibrium region, and does it even better, than Mao EOS, for instance.

Finalizing this section, though the proposed improved Murnaghan EOS was shown to be applicable for the successful fitting both of computational and experimental data, we have to point out here the weak side of the proposed two-step fitting process. It is high sensitivity to the quality of the treated sampling in the near-equilibrium region. If the latter contains too few points (or they are distributed too unevenly) there, the robustness of determination of $B_0$, $Bp$, $V_0$ parameters, and of the further fitting may be questionable. Though the computational sampling, in principle, may be obtained to be as dense as required, the available experimental data may be limited. However, the scheme of reliable obtaining of values of $B_0$, $Bp$, $V_0$ parameters from the data, related to the high-pressure region only, remains the open question yet.

# 5 The discussion

*(This section can be skipped on first reading)*

In the present section we shall discuss some more aspects of derivation and application of the improved Murnaghan EOS, which could be of interest, but were skipped yet for clarity and concision of presentation. In advance, though most likely there are no principal reasons, which would constrain the sign of $Bp$ parameter of solids (as well as the sign of *Bpp* and higher derivatives), the overwhelming majority of known crystalline compounds is characterized with its positive value. Although some exceptions are known (for example, according to [56], for the monoclinic $ZrO_2$ modification, the value of *Bp* was experimentally found to be negative (–3.6), in [57] the value (–15.88) was theoretically predicted for $Mn_3InN$ antiperovskite), we shall leave these "anomalies" beyond the discussion, for briefness, and shall assume further *Bp* to be always positive.

The first aspect to be mentioned is the behavior of *P*(V) EOS above the equilibrium point $V_0$. According to the "*ideal model*" of *P*(V) EOS behavior, the pressure should be positive below $V_0$, decrease with V growth, taking zero value at $V=V_0$, then take its minimum at *U*(V) flexion point, and further asymptotically tend to zero, as $V\to\infty$. But in fact EOSs are derived within some suggestions, and they may miss some aspects of the required *P*(V) behavior. As a result, some "unphysical" features of some EOSs may occur not too far from $V=V_0$ point, in particular, the EOS may predict zero pressure at some volume values besides $V_0$, which result in additional (and physically meaningless) extremes of the corresponding *U*(V) dependence. So, for third order Birch-Murnaghan EOS (2), it is not difficult to find, that it predicts the additional zero of pressure at volume value:

$$V_x = V_0\cdot[(3Bp–12)/(3Bp–16)]^{3/2}$$

if $Bp > 16/3$ or $Bp < 4$. While *Bp* value is not beyond ~6–7 (for a lot of typical compounds it is so), is seems not to be a problem, because $V_x>2.4\cdot V_0$, i.e. lies far enough above the equilibrium point and affects the EOS behavior in its vicinity a little. However, as *Bp* becomes ~10–12 (for $Mn_3CuN$ antiperovskite, for instance, *Bp* value was predicted to be 12.45 [57]), $V_x$ approaches closer to $V_0$ (~1.5–1.3 of its value), making EOS to deviate from its physically reasonable behavior. Moreover, at $Bp < 4$ zero of pressure is predicted even below $V_0$, for instance, at $Bp$=2.64 (predicted for $BaVO_3$ [17]) $V_x \approx 0.36\cdot V_0$, limiting the high-pressure region, where *P*(V) behavior is described by EOS (2) reliably. The similar situation also takes place for fourth order Birch-Murnaghan EOS (8), though its regarding becomes more complicated. Next, the third order logarithmic EOS (10) also predicts the additional zero of pressure, as volume takes value:

$$V_x = V_0 \cdot exp\{2/(Bp-2)\}.$$

It would be worth to mention also in the present context, that some EOSs may have the limitations on a region of their applicability, which, though, reveal only at some uncommon values of parameters. So, the region of applicability of Mao EOS (in its original form (15) [29], at least) is limited by the condition:

$$\left(V_0 / V\right)^{B_0 \cdot Bpp / Bp} \le e.$$

Since typically it is taken, that $Bp > 0$ and $Bpp < 0$, this condition is satisfied at $V<V_0$ (i.e. at hydrostatic compressions). However, otherwise it imposes the limitation on the available region of applicability. Such kind of situation encountered during fitting experimental V–$P$ data for $Zn_2SnO_4$, see Fig.5 and Table 2. The optimal value of b parameter of EOS (15) was found to be (–1.36), and, since $b = 1 - (B_0 \cdot Bpp/Bp^2)$ and $Bp>0$, it means that $Bpp > 0$, making the specified limitation to be active.

We shall offer to readers to search for further examples of vulnerable situations independently. Returning to our improved Murnaghan EOS (26), though it is difficult to propose the direct expression for its roots at zero pressure, besides $V=V_0$, it is possible, analyzing the behavior of first, second and third orders derivatives of (26), to find out that: 1) While $\alpha < 1$, the $P$(V) EOS (26) has neither zeroes nor extrema at $V < V_0$ and monotonously decreases, as V increases; If $0 < \alpha < 1$, it has a minimum at $V_m > V_0$, determined from the equation:

$$(V_0/V_m)^{Bp+1} = (\alpha/(\alpha-1)) \cdot \{(Bp+1) \cdot ((V_m/V_0)-1)-1\},$$

(which, though, is not able to be solved analytically too, but always has a solution) and provides zero of pressure at some $V>V_m$ value; If $\alpha < 0$, $P$(V) dependence (26) just monotonously decreases as V grows, having no zero values, besides at $V=V_0$. 2) While $\alpha > 1$, the $P$(V) EOS (26) has a maximum at volume value, lower than $V_0$, and zero of pressure below it, as well as a minimum at some $V > V_0$, and zero of pressure above it. Till now we did not confine neither sign nor magnitude of $\alpha$ parameter (though the requirement of negative *Bpp* value within PEM in geophysics results in demand $\alpha > 0$), at least, from the general solid state theory considerations. For the most of considered examples, both computational and experimental, the obtained $\alpha$ values fall into $0 < \alpha < 1$ range (though, for $\alpha$-Ca its value was rather close to unity), except $Zn_2SnO_4$, for which the negative $\alpha$ value (–4.375) was predicted. So, the latter condition ensures the correct behavior of EOS (26) at $V < V_0$ (i.e. in the practically meaningful region) for all the compounds, regarded in present work, however, it results in "unphysical" zero of pressure at some $V_x$ value above equilibrium value of volume. So, for $\alpha$-Ca, TiN, $SrTiO_3$, MgO, Si and diamond, the numerically calculated $V_x/V_0$ ratios are 1.37, 1.41, 1.58, 1.63, 1.70 and 2.01, respectively. For $\alpha$-Ca and titanium nitride, the extra "unphysical" zero of pressure appears not too far from the vicinity of $V_0$, corresponding to additional extremum ("unphysical" too) of *U*(V) dependence (28), such kind of its behavior explains the observed deviation in fitting the computational data for these crystalline solids above $V_0$ (see Fig.2). This way, though the modified Murnaghan EOS undoubtedly has serious advantages in fitting of computational *U*(V) data, as compared with ordinary EOS (13), it reveals, however, the hidden vulnerability, discussed above. It is unlikely, in author opinion, that the latter would complicate seriously the practical use of the modified EOS, at least, in the region of positive pressure values, however, it is desirable to keep the pointed out limitation in mind.

The next aspect we shall consider here, is the alternative approach to construct and derive the improved Murnaghan EOS, though, in slightly different form. Till now, it was skipped not to overburden our theoretical considerations, and to come sooner to their practical application. However, this alternative way, perhaps, also worth particular attention from the viewpoint of concepts of EOS construction. So, we shall return to the general form (16) of $P(\delta)$ expansion, which can be re-written in the form:

$$-B_0 \cdot Bp \cdot \delta = Bp \cdot P - (B_0 \cdot Bp/2) \cdot (Bp+1) \cdot \delta^2 - Bp \cdot p_3 \cdot \delta^3 - Bp \cdot p_4 \cdot \delta^4 - Bp \cdot p_5 \cdot \delta^5 - \ldots \quad (31)$$

(multiplying both sides by $Bp$), hence, the general form (17) of $B(\delta)$ expansion can be represented as follows:

$$B = B_0 + Bp \cdot P - [(B_0/2)(Bp+1)(Bp+2) + 3p_3] \cdot \delta^2 - [(3+Bp)p_3 + 4p_4] \cdot \delta^3 - \\ -[(4+Bp)p_4 + 5p_5] \cdot \delta^4 - [(5+Bp)p_5 + 6p_6] \cdot \delta^5 - \ldots - [(\mathrm{i}+Bp)p_\mathrm{i} + (\mathrm{i}+1)p_{\mathrm{i}+1}] \cdot \delta^\mathrm{i} - \ldots \quad (32)$$

On the other hand, the expansion (16) can be represented as:

$$P + B_0 \cdot \delta = (B_0/2) \cdot (Bp+1) \cdot \delta^2 + p_3 \cdot \delta^3 + p_4 \cdot \delta^4 + p_5 \cdot \delta^5 + p_6 \cdot \delta^6 + \ldots$$

Introducing the parameter $\gamma$, such as:

$$\gamma = Bp + 2 + 6 \cdot p_3/(B_0(Bp+1)), \quad (33)$$

and multiplying both sides of the previous relation by it, we obtain:

$$\gamma \cdot (P + B_0 \cdot \delta) = [(B_0/2)(Bp+1)(Bp+2) + 3 \cdot p_3] \cdot \delta^2 + \gamma \cdot \{p_3 \cdot \delta^3 + p_4 \cdot \delta^4 + p_5 \cdot \delta^5 + p_6 \cdot \delta^6 + \ldots\}.$$

Expressing the term $[\ldots] \cdot \delta^2$ in the explicit way from there and substituting it into (32), we can obtain after some rearrangement:

$$B = B_0 + (Bp - \gamma) \cdot P - B_0 \cdot \gamma \cdot \delta - [(3 + Bp - \gamma)p_3 + 4p_4] \cdot \delta^3 - [(4 + Bp - \gamma)p_4 + 5p_5] \cdot \delta^4 - \\ -[(5 + Bp - \gamma)p_5 + 6p_6] \cdot \delta^5 - \ldots - [(\mathrm{i} + Bp - \gamma)p_\mathrm{i} + (\mathrm{i}+1)p_{\mathrm{i}+1}] \cdot \delta^\mathrm{i} - \ldots \quad (34)$$

It should be noted, that the expansion of bulk modulus in form (34), unlike that in form (22), contains the term, linear in $\delta$, but does not contain the quadratic one, and it is its main feature. Substituting there $B = -(1+\delta) \cdot (dP(\delta)/d\delta)$ and neglecting the terms $\sim\delta^3$ and higher orders, we are able to solve the obtained differential equation in the same way as above, at initial condition $P(\delta=0)=0$, and finally to obtain:

$$P(\mathrm{V}) = \frac{B_0}{\tau+1} \cdot \left\{ \frac{(Bp+1)}{\tau} \cdot \left[ \left( \frac{\mathrm{V}_0}{\mathrm{V}} \right)^\tau - 1 \right] + (Bp - \tau) \cdot \left( \frac{\mathrm{V}}{\mathrm{V}_0} - 1 \right) \right\}, \quad (35)$$

where the new parameter $\tau = (Bp - \gamma)$ was introduced, for convenience. The latter equation is the alternative form of the improved Murnaghan EOS, somewhat different from (26), though including the same quantity of fitting parameters. Obviously, if $\gamma$ value is taken to be zero, the EOS (35) turns into ordinary Murnaghan EOS (1). This proposed form of EOS is more compact than (26), however, the procedure of two-steps fitting, described above, becomes more complicated and less transparent for it – due to appearance of term $(\mathrm{V}_0/\mathrm{V})^\tau$ therein instead of $(\mathrm{V}_0/\mathrm{V})^{Bp}$. However, the fitting of the computational $\{V_i, U_i\}$ data for the representative selection of compounds, regarded above, did not reveal yet any serious advantages of the EOS (35) in its corresponding $U$(V) form (the author offers to the readers to obtain it independently), as compared with (28). So, we just have derived it to illustrate the possible variation of the approach to improve the Murnaghan EOS, proposed here.

Finally, we shall discuss here the aspect, briefly mentioned in "The theoretical overview" section and concerning the choice of the small parameter, which would characterize the strain of the system. Up to now, we have been employing the relative compression $\delta = (V–V_0)/V_0$ in our considerations, for simplicity and obviousness. However, it is not the only choice, moreover, it is not proven to be optimal for the convergence of the different series, mentioned above. In fact, introducing the small parameter $f$ of more general form, $f = (V/V_0)^{n/3}–1$ (at n=2 and –2 it corresponds to the Lagrangian and Eulerian strains (6), respectively), we can see from Fig.6, that its squared (for more obviousness) values at n=±1, 2 are lower, than those at n=3, and increase slower, as the deviation of $V/V_0$ ratio from unity grows, – in whole region of reasonable $V/V_0$ variations (except at n=–2, where $V/V_0 < 0.6$). This way, the small parameter $f$, taken with n=±1, 2, can be hoped to provide more efficient convergence of the series, we deal above, at not too small strains, than $f = V/V_0–1$ is able. Moreover, the expansion (5) of free Helmholtz energy allows various forms of $f = f(V)$ dependence, like $f = ln(V/V_0)$, $[exp(V/V_0)–1]$, etc. And our proposed way to construct the improved Murnaghan EOS should not necessarily be built on $\delta = (V–V_0)/V_0$ as a small parameter. By analogy with the general form of $P(\delta)$ expansion (16), the expansion of pressure in terms of arbitrary small parameter $f$ also can be regarded:

$$P(f) = \pi_1{\cdot}f + \pi_2{\cdot}f^2 + \pi_3{\cdot}f^3\ \pi_4{\cdot}f^4 + \ldots + \pi_k{\cdot}f^k, \text{ where } \pi_k = (d^kP/df^k)|_{f=0}/k!$$

For substantive considerations, let us choose $f$ to be the same, as in derivations of Birch-Murnaghan EOS (2), (7), (8), i.e, as the Eulerian finite strain with inverse sign, $f = [(V/V_0)^{-2/3}–1]/2$ (with other choice, all further considerations are easily reproduced by analogy). Then it is straightforward to obtain:

$$P(f) = 3B_0{\cdot}f + (3B_0/2)(3Bp–2){\cdot}f^2 + \pi_3{\cdot}f^3\ \pi_4{\cdot}f^4 + \ldots \quad (36)$$

And the general form of bulk modulus expansion, analogous to (17), is obtained to have a form:

$$B(f) = B_0 + 3B_0Bp\cdot f + [2B_0(3Bp–2)+\pi_3]\cdot f^2 + [2\pi_3+4\pi_4/3]\cdot f^3 + \ldots \quad (37)$$

The assumption (12), underlying the Murnaghan EOS, in terms of (36) can be represented as:

$$B = B_0 + 3B_0Bp\cdot f + (3B_0/2)Bp(3Bp–2)\cdot f^2 + \pi_3\cdot Bp\cdot f^3 + \pi_4\cdot Bp\cdot f^4 + \ldots \quad (38)$$

Further, we do the same, we did in (21), i.e. subtract from (12) the "*incorrect*" form of the term, proportional to $f^2$, taken from (38), and add the "*right*" one, taken from (37). Neglecting the terms, $\sim f^3$ and higher orders, we now have:

$$B = B_0 + Bp\cdot P + [\pi_3 – (B_0/2)(3Bp–2)(3Bp–4)]\cdot f^2 \quad (39)$$

Taking into account, that $B = (1+2\cdot f)\cdot(\mathrm{d}P/\mathrm{d}f)/3$, we can solve the obtained differential equation with the variation of constants method at $P(f=0)=0$ condition. The author offers again to exercise in math independently, just providing the final result:

(40)

$$P(\mathrm{V}) = \frac{B_0}{Bp}\cdot(1-\alpha)\cdot\left[\left(\frac{\mathrm{V}_0}{\mathrm{V}}\right)^{Bp} - 1\right] + \frac{3\alpha\cdot B_0}{4}\cdot\left[\left(\frac{\mathrm{V}}{\mathrm{V}_0}\right)^{-2/3} - 1\right]\cdot\left\{2 + \left(\frac{3Bp}{2} - 1\right)\cdot\left[\left(\frac{\mathrm{V}}{\mathrm{V}_0}\right)^{-2/3} - 1\right]\right\},$$

where

$$\alpha = 1 - \frac{2\cdot\pi_3}{B_0\cdot(3Bp-2)(3Bp-4)} \quad .$$

(note, that the α parameter, introduced in (26) is not quantitatively identical to that, introduced in (40), during the derivation of EOSs they were determined in a different way, though the structure of the obtained equation is quite similar). However, the tests of the obtained EOS (40) by fitting the computational data, as well as the searches of possibly more efficient forms of small parameter $f$, are left beyond the present paper – with the modest hope to motivate interest of the readers to such kind of matter. The possibility of different choice of $f$ parameter may provide the additional "*hidden flexibility*" of the constructed EOS and it is worth to be taken into account too.

# 6 The conclusion

In summary, we have overviewed some of the known EOSs for solids and regarded the concepts and approaches, underlying the algorithms to derive them. Based on the mentioned aspects, the weak side, common for a lot of them (including the "classical" Murnaghan EOS), has been pointed out – the scarcity of "fitting flexibility". EOSs of third order are unable to reproduce equally accurately both the near-equilibrium and high-pressure regions simultaneously (besides the particular "happy" examples). Generally, at formal overall fitting the good accordance to data within one of mentioned regions would result in poor agreement within another one, and *vice versa*. For Murnaghan EOS, its improved form, obtained as a result of the simple "correction" of the underlying empirical relation $B = B_0 + Bp{\cdot}P$, has been proposed. It is not too complicated in comparison with original EOS and includes, besides standard fitting parameters $B_0$, $Bp$ and $V_0$, one more parameter, responsible for high-pressure behaviour of EOS and providing additional "flexibility" to it. This improved EOS has shown its advantages in comparison with the ordinary one, during the fitting the computational $U$(V) data for the representative series of soilds: α-Ca, MgO, TiN, diamond, silicon and $SrTiO_3$, since it allows to broaden the region of reliable data reproducing considerably, up to pressures, higher than $B_0$ value (for ordinary Murnaghan EOS, the limitation $\sim B_0/2$ typically is taken). It also has shown its applicability to treat the results of experimental V–$P$ measurements, fitting, as an example, the data, available for metallic chromium, ruthenium and $Zn_2SnO_4$ stannate – in comparison both with widely used Vinet EOS and equations of fourth order: Birch-Murnaghan and Mao EOS. For $Zn_2SnO_4$, where the V–$P$ data sampling turned out to be "difficult" for Vinet EOS to fit it as a whole, the proposed equation has reproduced it reasonably both in near-equilibrium and high-pressure regions, at least, better, than Mao EOS do. The two-steps fitting process was proposed to determine the parameters $B_0$, $Bp$ and $V_0$ of EOS accurately. It implies the "binding" of EOS to the near-equilibrium region at the first stage and is based on the principle, that $B_0$, $Bp$ and $V_0$, determining the low-pressure behavior of EOS, can and should be obtained first from the data, related to near-equilibrium region (where the variation of other

EOS parameters does insignificant effect). Some additional aspects, such as the hidden vulnerability of the improved EOS, resulting in possible "unphysical" behavior of pressure at $V>V_0$ (not very important for practical applications, though) and alternative approaches, allowing to obtain the improved Murnaghan EOS in somewhat different form, also were discussed. The author modestly hopes to motivate the interest of the readers to the problem of developing and improvement the EOSs for solids, because the various approximations, underlying the known equations, are resulting in the limitations and difficulties for their practical applicability from time to time.

## Acknowledgement

The author is sincerely thankful to Ludmila N. Bannikova for her help and irreplaceable support during the work and writing process.

## Declaration of competing interest

The author declares that he has no known competing financial interests or personal relationships that could have appeared to influence the work reported in this paper.

## Supplementary materials

The Supplementary materials can be downloaded from the Web page:
*https://www.researchgate.net/publication/414045883_Supplementary_materials_-_How_to_enhance_the_applicability_of_Murnaghan_equation_of_state*
or can be provided on reasonable request via Email bannikov@ihim.uran.ru

## References

[1] D. Laniel, F. Trybel, B. Winkler, F. Knoop, T. Fedotenko, S. Khandarkhaeva, A. Aslandukova, T. Meier, S. Chariton, K. Glazyrin, V. Milman, V. Prakapenka, I.A.

Abrikosov, L. Dubrovinsky, N. Dubrovinskaia // Nature Communications., **13**, 6987 (2022). DOI: 10.1038/s41467-022-34755-y

[2] E. Gregoryanz, C. Sanloup, M. Somayazulu, J. Badro, G. Fiquet, H.-K. Mao, R.J. Hemley // Nature Materials, **3**, p.294 (2004). DOI: 10.1038/nmat1115

[3] S. Ono, T. Kikegawa, Y. Ohishi // Sol. St. Comm., **133**, p.55 (2005). DOI: 10.1016/j.ssc.2004.09.048

[4] K. Nishimura, I. Yamada, K. Oka, Y. Shimakawa, M. Azuma // J. Phys. Chem. Sol., **75**, p.710 (2014). DOI: 10.1016/j.jpcs.2014.02.001

[5] O.L. Anderson. Equations of State of Solids for Geophysics and Ceramic Science. // N.Y., Oxford University Press, 1995. ISBN 0-19-505606-X. DOI: 10.1093/oso/9780195056068.001.0001

[6] J.R. Macdonald // Rev. Mod. Phys., **41** (№2), p.316 (1969). DOI: 10.1103/RevModPhys.41.316

[7] F.D. Stacey, B.J. Brennan, R.D. Irvine // Geophys. Surv., **4**, p.189 (1981). DOI: 10.1007/BF01449185

[8] F.D. Murnaghan // Proc. N.A.S., **30** (№9), p.244 (1944). DOI: 10.1073/pnas.30.9.244

[9] F. Birch // Phys. Rev., **71** (№11), p.809 (1947). DOI: 10.1103/PhysRev.71.809

[10] P. Vinet, J. Ferrante, J.R. Smith, J.H. Rose // J. Phys. C: Sol. St. Phys., **19** (№20), L467 (1986). DOI: 10.1088/0022-3719/19/20/001

[11] P. Vinet, J.R. Smith, J. Ferrante, J.H. Rose // Phys. Rev. B, **35** (№4), p.1945 (1987). DOI: 10.1103/PhysRevB.35.1945

[12] H. Anton, P.C. Schmidt // Intermetallics, **5**, p.449 (1997). DOI: 10.1016/S0966-9795(97)00017-4

[13] B. Mayer, H. Anton, E. Botta, M. Methfessel, J. Sticht, J. Harris, P.C. Schmidt // Intermetallics, **11**, p.23 (2003). DOI: 10.1016/S0966-9795(02)00127-9

[14] T. Katsura, Y. Tange // Minerals, **9**, 745 (2019). DOI: 10.3390/min9120745

[15] J.-P. Poirier, A. Tarantola // Phys. Earth Planet. Inter., **109**, p.1 (1998). DOI: 10.1016/S0031-9201(98)00112-5

[16] M. E. Nasab, S.S. Tafreshi, S.Jouybar, L. Naji // Mater. Chem. Phys., **344**, 131085 (2025). DOI: 10.1016/j.matchemphys.2025.131085

[17] V.V. Bannikov // Mater. Chem. Phys., **171**, p.119 (2016). DOI: 10.1016/j.matchemphys.2015.12.007

[18] U. Walzer // Phys. Earth Planet. Inter., **30**, p.62 (1982). DOI: 10.1016/0031-9201(82)90128-5

[19] S.P. Singh, S. Gupta, S.C. Goyal // Physica B, **391**, p.307 (2007). DOI: 10.1016/j.physb.2006.10.011

[20] V.G. Tyuterev, N. Vast // Comp. Mat. Sci., **38**, p.350 (2006). DOI: 10.1016/j.commatsci.2005.08.012

[21] R. Grover, I.C. Getting, G.C. Kennedy // Phys. Rev. B, **7** (№2), p.567 (1973). DOI: 10.1103/PhysRevB.7.567

[22] A.M. Dziewonski, A.L. Hales, E.R. Lapwood // Phys. Earth Planet. Inter., **10**, p.12 (1975). DOI: 10.1016/0031-9201(75)90017-5

[23] K. Fuchizaki // J. Phys. Soc. Jap., **75** (№3), 034601 (2006). DOI: 10.1143/JPSJ.75.034601

[24] T.A. Wani, B.K. Das, B. Tripathi, I.A. Khan // In: Singh Tomar, G., Chaudhari, N.S., Barbosa, J.L.V., Aghwariya, M.K. (eds) International Conference on Intelligent Computing and Smart Communication 2019. Algorithms for Intelligent Systems. Springer, Singapore, p.841 (2020). DOI: 10.1007/978-981-15-0633-8_87

[25] J.R. Macdonald // Rev. Mod. Phys., **38** (№4), p.669 (1966). DOI: 10.1103/RevModPhys.38.669

[26] S.K. Saxena // J. Phys. Chem. Solids, **65**, p.1561 (2004). DOI: 10.1016/j.jpcs.2004.02.003

[27] K. Anand, M.P. Singh, B.S. Sharma // J. Phys. Chem. Solids, **134**, p.121 (2019). DOI: 10.1016/j.jpcs.2019.05.039

[28] K. Rajesh, K.S. Singh, K. Anand // Comp. Cond. Matter, **32**, e00710 (2022). DOI: 10.1016/j.cocom.2022.e00710

[29] N.-H. Mao // J. Geophys. Research, **75** (№35), p.7508 (1970). DOI: 10.1029/JB075i035p07508

[30] S. Anzellini, D. Diaz-Anichtchenko, J. Sanchez-Martin, R. Turnbull, S. Radescu, A. Mujica, A. Munoz, S. Ferrari, L. Pampillo, V. Bilovol, C. Popescu, D. Errandonea // J. Phys. Chem. C, **128**, p.1357 (2024). DOI: 10.1021/acs.jpcc.3c06726

[31] S. Anzellini, D. Errandonea, L. Burakovsky, J.E. Proctor, R. Turnbull, C.M. Beavers // Sci. Rep., **12**:6727 (2022). DOI: 10.1038/s41598-022-10523-2

[32] S. Anzellini, D. Errandonea, C. Cazorla, S. MacLeod, V. Monteseguro, S. Boccato, E. Bandiello, D. D. Diaz Anichtchenko, C. Popescu, C.M. Beavers // Sci. Rep., **9**:14459 (2019). DOI: 10.1038/s41598-019-51037-8

[33] D.J Singh, L. Nordstrom // Planewaves, Pseudopotentials and the LAPW method (2nd edition). Boston, MA: Springer (2006). DOI: 10.1007/978-0-387-29684-5

[34] J.P. Perdew, S. Burke, M. Ernzerhof // Phys. Rev. Lett., **77**, p.3865 (1996). DOI: 10.1103/PhysRevLett.77.3865

[35] P. Blaha, K. Schwarz, G.K.H. Madsen, D. Kvasnicka, J. Luitz. // WIEN2k, An Augmented Plane Wave Plus Local Orbitals Program for Calculating Crystal Properties. Vienna: Vienna University of Technology (2009).

[36] Q.F. Gu, G. Krauss, Yu. Grin, W. Steurer // Phys. Rev. B, **79**, 134121 (2009). DOI: 10.1103/PhysRevB.79.134121

[37] S. Anzellini, D. Errandonea, S.G. MacLeod, P. Botella, D. Daisenberger, J.M. De`Ath, J. Gonzalez-Platas, J. Ibanez, M.I. McMahon, K.A. Munro, C. Popescu, J. Ruiz-Fuertes, C.W. Wilson // Phys. Rev. Mater., **2**, 083608 (2018). DOI: 10.1103/PhysRevMaterials.2.083608

[38] V.P. Itkin, C.B. Alcock // J. Phase Equil., **11**(№5), p. 497 (1990). DOI: 10.1007/BF02898268

[39] R.C. Ropp // Encyclopedia of the Alkaline Earth Compounds. p. 15 // Elsevier (2013), ISBN: 978-0-444-59550-8. DOI: 10.1016/B978-0-444-59550-8.00001-6

[40] S.D. Jacobsen, C.M. Holl, K.A. Adams, R.A. Fischer, E.S. Martin, C.R. Bina, J-F. Lin, V.B. Prakapenka, A. Kubo, P. Dera // Amer. Mineral., **93**, p.1823 (2008). DOI: 10.2138/am.2008.2988

[41] Y. Kono, T. Irifune, Y. Higo, T. Inoue, A. Barnhoorn // Phys. Earth Planet. Inter., **183**, p.196 (2010). DOI: 10.1016/j.pepi.2010.03.010

[42] A. Dewaele, G. Fiquet // J. Geophys. Research, **105** (№B2), p.2869 (2000). DOI: 10.1029/1999JB900364

[43] S. Yoon, A.E. Maegli, L. Karvonen, S.K. Matam, A. Shkabko, S. Riegg, T. Grosmann, S.G. Ebbinghaus, S. Pokrant, A. Weidenkaff // J. Sol. St. Chem., **206**, p.226 (2013). DOI: 10.1016/j.jssc.2013.08.001

[44] A.K. Singh, G.C. Kennedy // J. Appl. Phys., **48**(№8), p.3362 (1977). DOI: 10.1063/1.324175

[45] H.J. McSkimin, P. Andreatch, Jr. // J. Appl. Phys., **35**(№7), p.2161 (1964). DOI: 10.1063/1.1702809

[46] H.J. McSkimin, P. Andreatch, Jr. // J. Appl. Phys., 4**3**(№7), p.2944 (1972). DOI: 10.1063/1.1661636

[47] M.H. Grimsditch, A.K. Ramdas // Phys. Rev. B, **11**(№8), p.3139 (1975). DOI: 10.1103/PhysRevB.11.3139

[48] D.J. Dunstan, S.H.B. Bosher // Phys. Stat. Sol. (b), **235**(№2), p.396 (2003). DOI: 10.1002/pssb.200301591

[49] Q. Wang, D. He, F. Peng, L. Xiong, J. Wang, P. Wang, C. Xu, J. Liu // Sol. St. Comm., **182**, p.26 (2014). DOI: 10.1016/j.ssc.2013.12.015

[50] H. Chen, F. Peng, H-k. Mao, G. Shen, H.-P. Liermann, Z. Li, J. Shu // J. Appl. Phys., **107**, 112503 (2010). DOI: 10.1063/1.3392848

[51] N.J. Perks, Z. Zhang, R.J. Harrison, M.A. Carpenter // J. Phys.: Cond. Matt., **26**, 505402 (2014). DOI: 10.1088/0953-8984/26/50/505402

[52] R.O. Bell, G. Rupprecht // Phys. Rev, **129**(№1), p.90 (1963). DOI: 10.1103/PhysRev.129.90

[53] A. Dewaele, M. Torrent, P. Loubeyre, M. Mezouar // Phys. Rev. B, **78**, 104102 (2008). DOI: 10.1103/PhysRevB.78.104102

[54] M.A. Mahammed, H.B. Mohammed // Adv. Cond. Matt. Phys., **2023**, 9518475 (2023). DOI: 10.1155/2023/9518475

[55] B.-W. Kim, C. Liu, H. Yin // J. Appl. Phys. **135**, 075105 (2024). DOI: 10.1063/5.0184120

[56] M. Fujimoto, Y. Akahama, H. Fukui, N. Hirao, Y. Ohishi // AIP Advances, **8**, 015310 (2018). DOI: 10.1063/1.5017774

[57] V.V. Bannikov // Comp. Cond. Matter, **32**, e00729 (2022). DOI: 10.1016/j.cocom.2022.e00729

Table 1. The values of the parameters $B_0$, $Bp$, $V_0$ and $\alpha$ (if it is present), obtained from the fitting of the computational $\{V_i, U_i\}$ samplings both with ordinary (13) and improved Murnaghan EOS (28). The results of fitting both of full and "shortened" samplings (see text) by EOS (28) are provided, for comparison.

| | | EOS (13) | EOS (28) – full * | EOS (28) – shortened |
|---|---|---|---|---|
| MgO | $B_0$, GPa | 146.8605 | 147.5964 | 147.4345 |
| | $Bp$ | 4.1113 | 4.1437 | 4.1375 |
| | $V_0$, bohr$^3$ | 130.5125 | 130.5084 | 130.5093 |
| | $\alpha$ | – | 0.3551 | 0.2871 |
| $\alpha$-Ca | $B_0$, GPa | 17.7504 | 18.2067 | 18.2803 |
| | $Bp$ | 4.6601 | 4.6962 | 4.6912 |
| | $V_0$, bohr$^3$ | 286.6412 | 286.6147 | 286.6139 |
| | $\alpha$ | – | 0.8955 | 1.0672 |
| Diamond | $B_0$, GPa | 440.2135 | 440.9642 | 440.8612 |
| | $Bp$ | 3.4149 | 3.4239 | 3.4227 |
| | $V_0$, bohr$^3$ | 76.9416 | 76.9409 | 76.9410 |
| | $\alpha$ | – | 0.1773 | 0.1698 |
| Silicon | $B_0$, GPa | 88.5314 | 88.7521 | 88.6933 |
| | $Bp$ | 4.2728 | 4.2164 | 4.2302 |
| | $V_0$, bohr$^3$ | 276.4988 | 276.5023 | 276.5014 |
| | $\alpha$ | – | 0.2673 | 0.2053 |
| TiN | $B_0$, GPa | 278.2519 | 281.5489 | 281.8874 |
| | $Bp$ | 5.0488 | 5.0161 | 5.0102 |
| | $V_0$, bohr$^3$ | 129.4217 | 129.4193 | 129.4193 |
| | $\alpha$ | – | 0.6193 | 0.7434 |
| $SrTiO_3$ | $B_0$, GPa | 169.2767 | 170.0715 | 170.0568 |
| | $Bp$ | 4.6084 | 4.5344 | 4.5358 |

| | | | |
|---|---|---|---|
| $V_0$, bohr$^3$ | 412.8096 | 412.8176 | 412.8175 |
| $\alpha$ | – | 0.3514 | 0.3704 |

* The data in this column are regarded as the final computational results, compared with available experiment. To make the results of various fitting schemes more convenient for comparison, the data are provided in the present form. The $V_0$ values are related to the corresponding *primitive* cells (containing two atoms for diamond and silicon, one atom for $\alpha$-Ca, and one formula unit for other compounds). For more obvious comparison with experimental data, the values of $V_0$ are provided in the text in Å$^3$ units and are related to the *conventional* cubic cell (eight atoms for diamond and silicon, four atoms for $\alpha$-Ca, four formula units for TiN and MgO, and one for $SrTiO_3$). In the text the values are rounded in correspondence with experimental ones.

Table 2. The optimal values of the EOSs parameters, obtained during the fitting of the specified experimental V–*P* data [29-31]. For BM4, the notation $\omega=[9\cdot B_0\cdot Bpp - 9\cdot Bp\cdot(7-Bp) + 143]/24$ is used, for briefness. The calculated values of RMSD for each fitting result also are specified.

| | $B_0$, GPa | *Bp* | $V_0$, Å$^3$ | Additional | RMSD, GPa |
|---|---|---|---|---|---|
| | | | Chromium | | |
| EOS (26) | 192.58 | 4.486 | 24.049 | $\alpha = 0.204$ | 0.555 |
| Vinet [31] | 182.0 | 5.10 | 24.08 | – | 0.563 |
| BM3 [31] | 185.0 | 4.74 | 24.08 | – | 0.498 |
| | | | Ruthenium | | |
| EOS (26) | 306.24 | 4.595 | 27.170 | $\alpha = 0.243$ | 0.691 |
| Vinet [32] | 319.1 | 4.40 | 27.129 | – | 0.467 |
| BM3 [32] | 323.4 | 4.15 | 27.122 | – | 0.467 |
| | | | $Zn_2SnO_4$ | | |
| EOS (26) | 151.26 | 5.480 | 649.185 | $\alpha = -4.375$ | 0.227 |
| Vinet [30] | 150.0 | 7 | 649.3 | – | 1.093 |
| BM4 | 153.51 | 4.188 | 649.153 | $\omega = 83.763$ | 0.204 |
| Mao | 153.35 | 4.856 | 649.148 | a = 31.578 GPa; b = –1.36 | 0.446 |
| EOS (27) | 151.27 | 6.893 | 649.170 | $\alpha = -15.991$; $\lambda = 199.781$ | 0.197 |

FIGURES

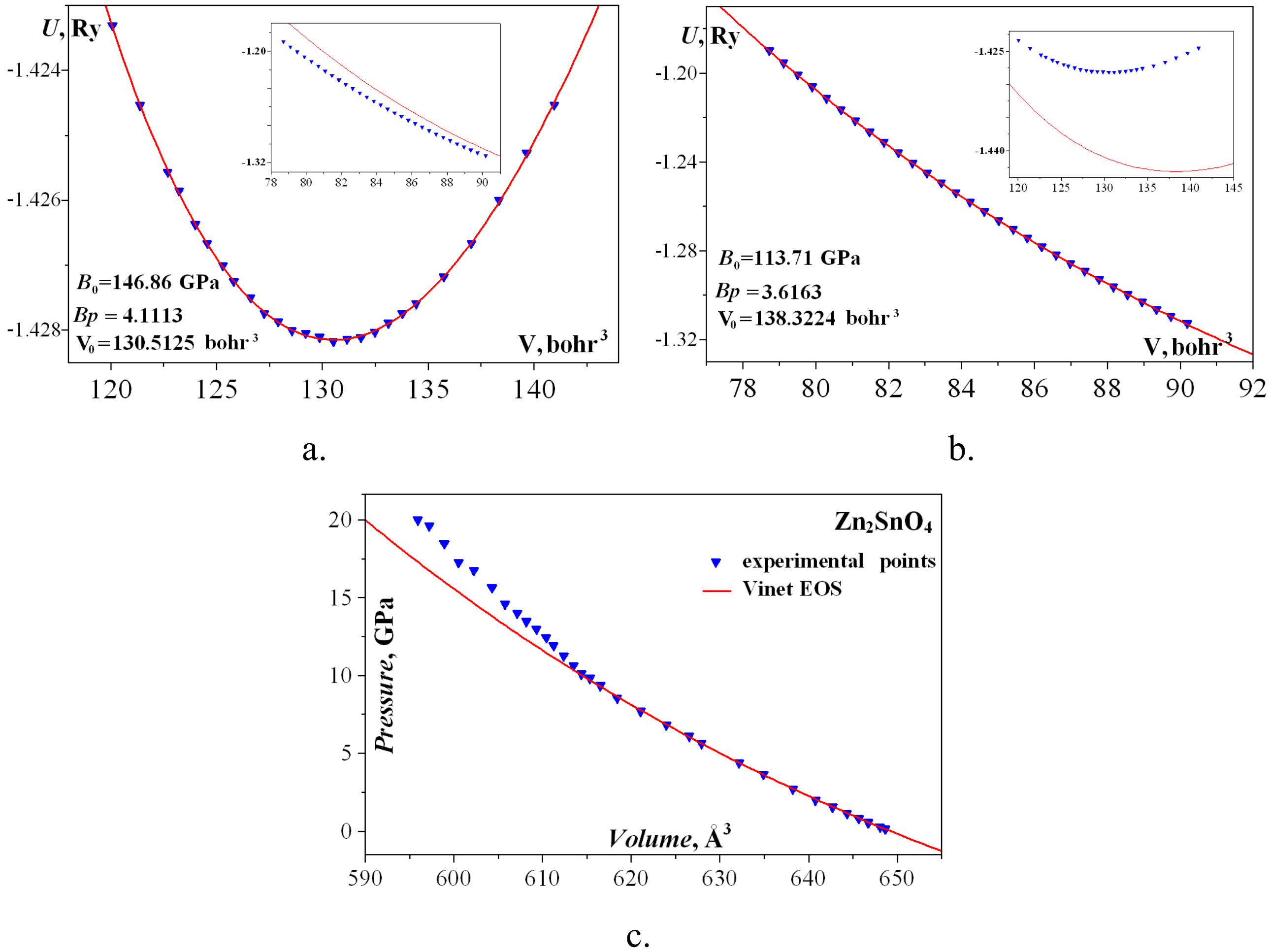

Fig. 1. The illustration of the problem of EOS "flexibility". (a) The results of separate fitting of $U$(V) computational data for MgO by EOS (13) in near-equilibrium region and (b) in high-pressure region. The insets show the quality of reproducing of "opposite" regions, the obtained values of $B_0$, $Bp$, $V_0$ also are provided (the values of $U_0$ are –1.4281 Ry and –1.4431 Ry for (a) and (b), respectively). These regions correspond to pressure values (a) up to ~14 GPa, and (b) ~115-210 GPa (estimated from EOS (26), see text). (c) The experimental V($P$) dependence up to 20 GPa, obtained for $Zn_2SnO_4$ stannate, and its fitting by Vinet EOS (the reproduced data of [30]).

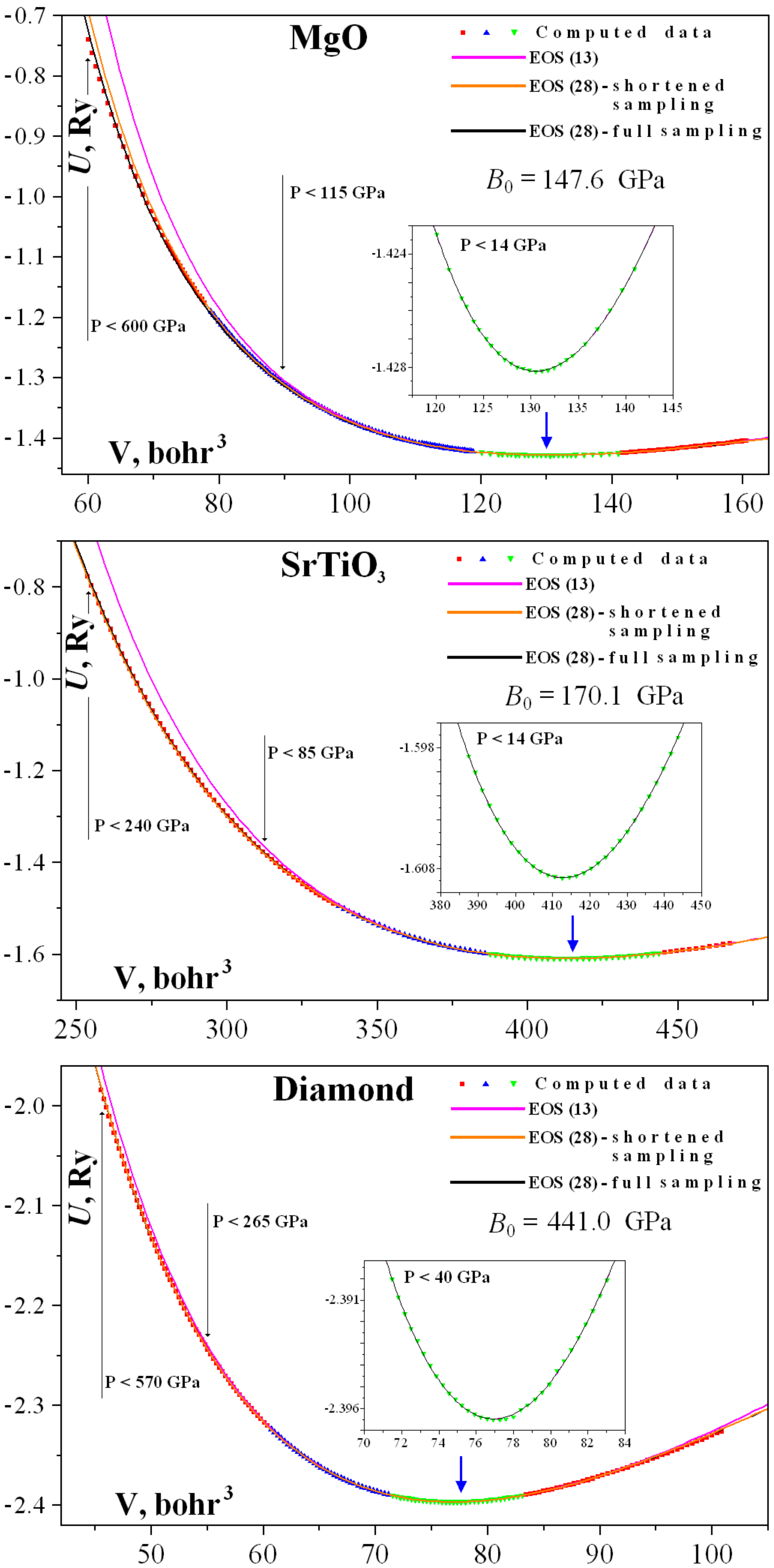
MgO
Computed data
EOS (13)
EOS (28) - shortened sampling
EOS (28) - full sampling
$B_0$ = 147.6 GPa
P < 115 GPa
P < 600 GPa
P < 14 GPa
U, Ry
V, bohr³
SrTiO₃
Computed data
EOS (13)
EOS (28) - shortened sampling
EOS (28) - full sampling
$B_0$ = 170.1 GPa
P < 85 GPa
P < 240 GPa
P < 14 GPa
U, Ry
V, bohr³
Diamond
Computed data
EOS (13)
EOS (28) - shortened sampling
EOS (28) - full sampling
$B_0$ = 441.0 GPa
P < 265 GPa
P < 570 GPa
P < 40 GPa
U, Ry
V, bohr³

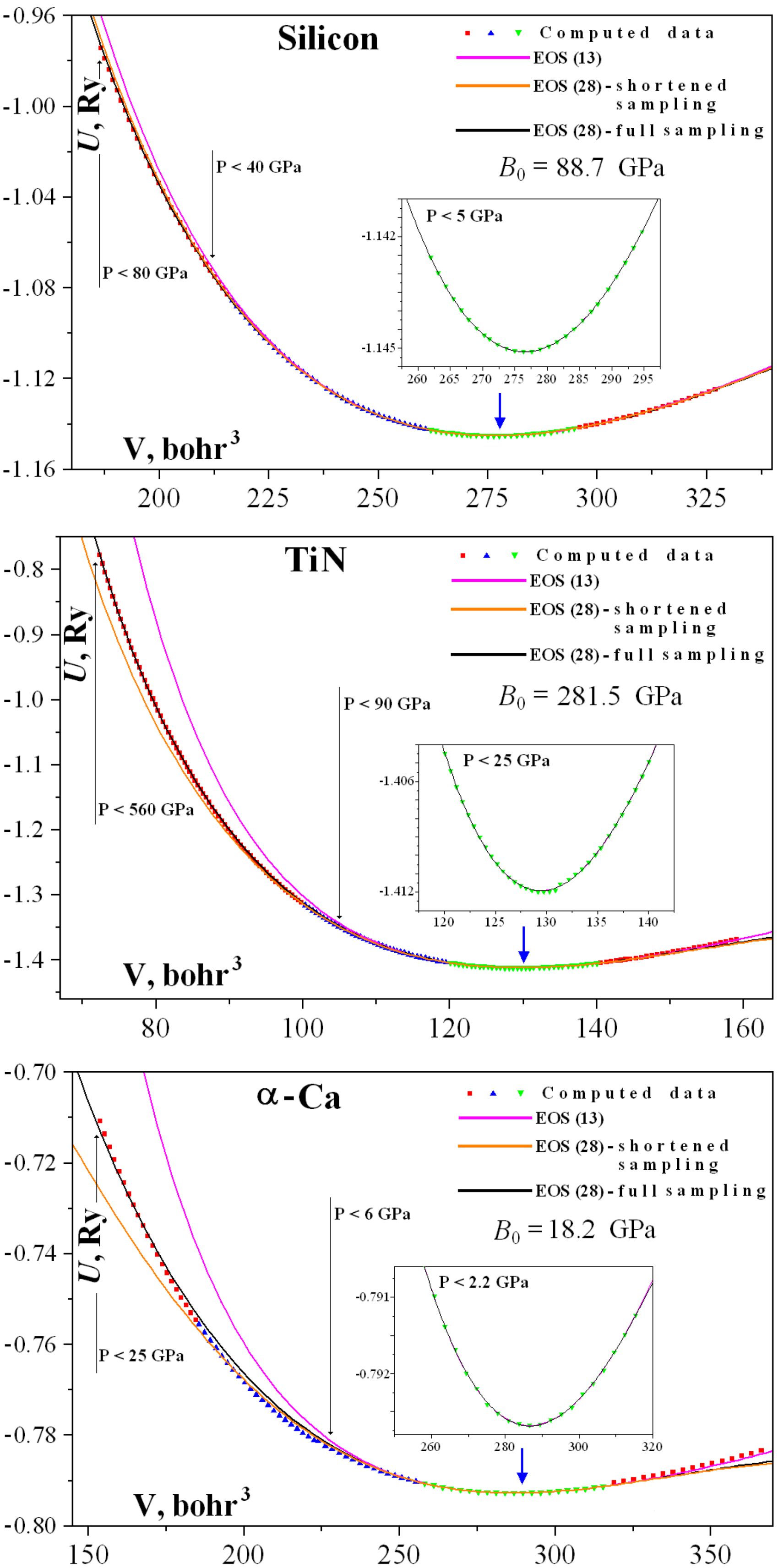

Silicon
Computed data
EOS (13)
EOS (28) - shortened sampling
EOS (28) - full sampling
B0 = 88.7 GPa
P < 40 GPa
P < 80 GPa
P < 5 GPa
U, Ry
V, bohr3
TiN
B0 = 281.5 GPa
P < 90 GPa
P < 560 GPa
P < 25 GPa
α-Ca
B0 = 18.2 GPa
P < 6 GPa
P < 25 GPa
P < 2.2 GPa

Fig. 2. The computational $U$(V) dependence (or $\{V_i, U_i\}$ sampling) for the representative selection of compounds: MgO, $SrTiO_3$, diamond, silicon, TiN, α-Ca, and its fitting both by ordinary Murnaghan EOS (13) and the improved EOS (28). For the latter, the fitting both of full and "shortened" (i.e. green and blue points) sampling was performed (see the Table 1 and text for details). The behavior of EOSs in near-equilibrium region is shown on insets for each compound.

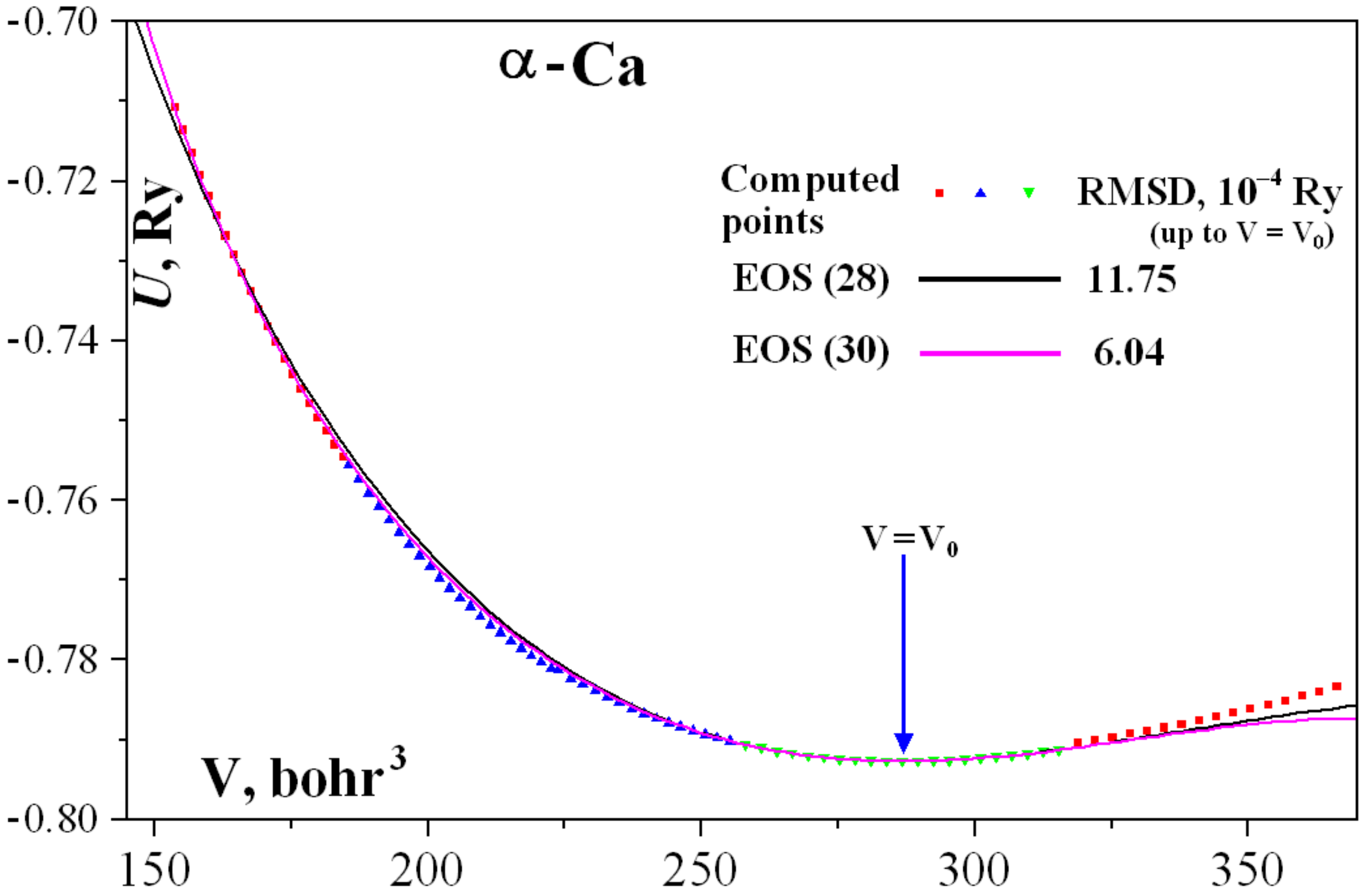

Fig. 3. The comparative fitting of $\{V_i, U_i\}$ sampling for α-Ca with improved Murnaghan EOS of second (28) and third (30) orders. The calculated values of RMSD (within the region of $V<V_0$) also are provided.

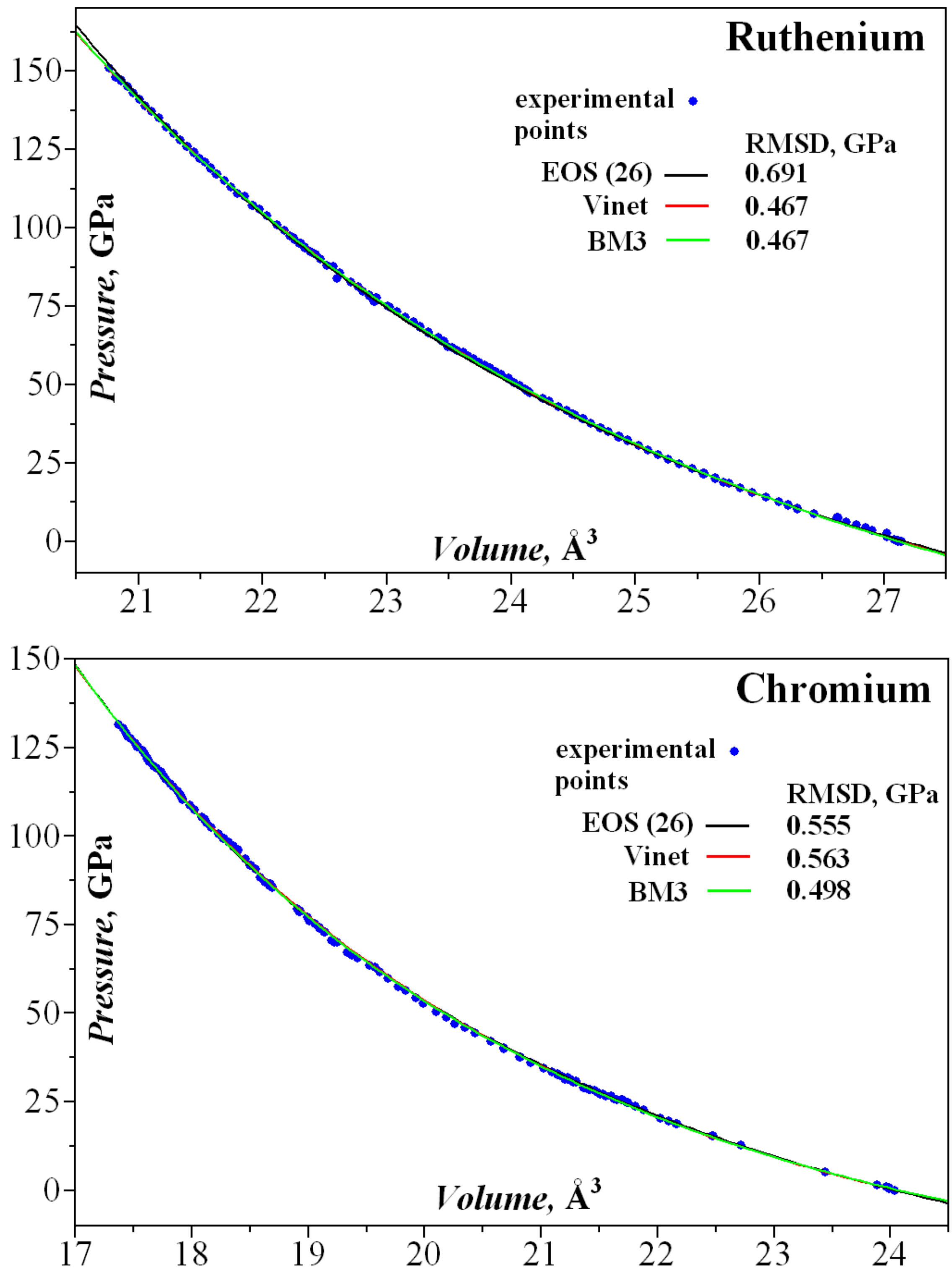

Fig. 4. The results of experimental V–*P* measurements for chromium [31] and ruthenium [32], and their fitting by improved Murnaghan EOS (26) – in comparison with Vinet (3) and Birch-Murnaghan (BM3) EOS (2). The calculated RMSD values also are specified in the plots (see also Table 2).

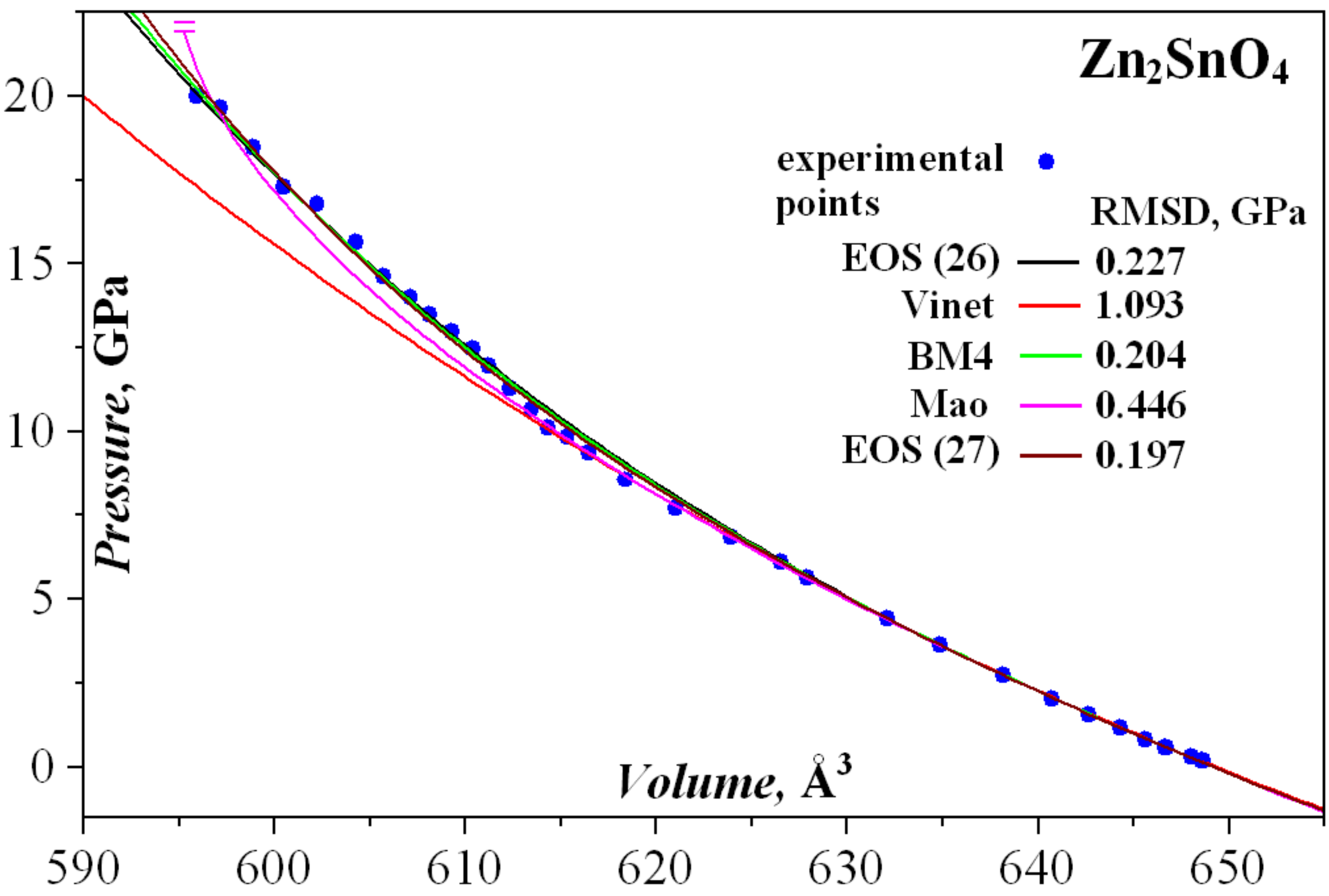

Fig. 5. The fitting of experimental V-*P* data sampling for $Zn_2SnO_4$ stannate [30] by improved Murnaghan EOS (26) and EOS (27) in comparison with BM4 (8) and Mao EOS (15). The behavior of the Vinet EOS with the parameters, optimized within the region up to 10 GPa, also is shown, for obviousness. The sign "=" (magenta plot) denotes the limitation of the region of Mao EOS applicability (see Discussion section).

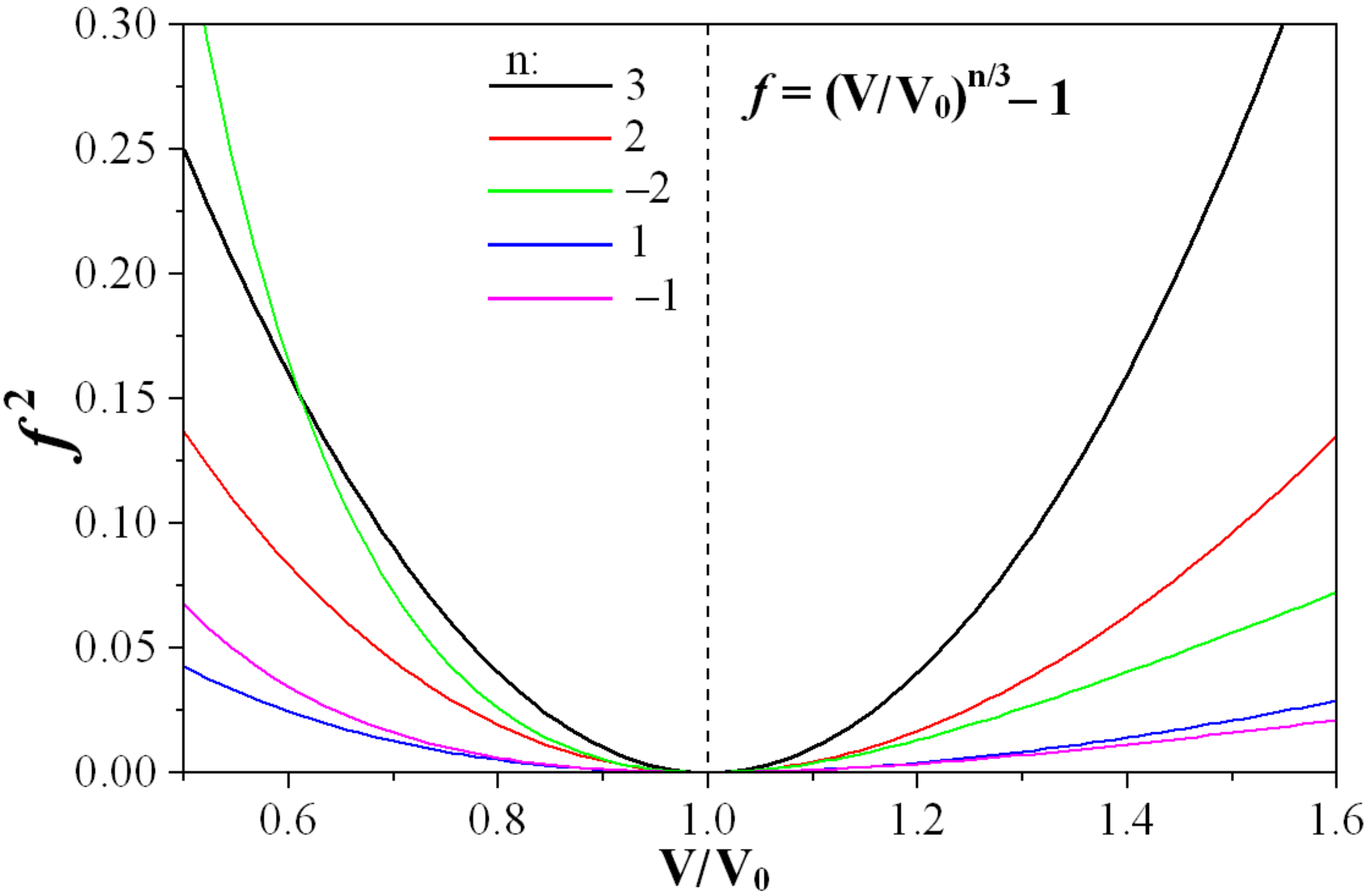

Fig.6. The behavior of the square of the function $f(V) = (V/V_0)^{n/3} - 1$ at various n values.